\documentclass[%
 reprint,
 amsmath,amssymb,
 aps,
]{revtex4-1}
\usepackage{silence}
\usepackage[utf8]{inputenc}
\usepackage{graphicx}
\usepackage{dcolumn}
\usepackage{bm}
\usepackage{ulem}
\usepackage{xcolor}
\usepackage{lipsum}
\usepackage{hyperref}
\hypersetup{
    colorlinks=true,
    linkcolor=blue,
    filecolor=magenta,      
    urlcolor=black,
    citecolor=blue,
    pdftitle={Overleaf Example},
    pdfpagemode=FullScreen,
    }

\begin{document}

\preprint{APS/123-QED}

\title{Quantum-like nonlinear interferometry with frequency-engineered classical light}

\author{Romain Dalidet, Anthony Martin, Grégory Sauder, Laurent Labonté}
\author{Sébastien Tanzilli}
\affiliation{Université Côte d’Azur, CNRS, Institut de physique de Nice, France}

\begin{abstract}
Quantum interferometry methods exploit quantum resources, such as photonic entanglement, to enhance phase estimation beyond classical limits. Nonlinear optics  has served as a workhorse for the generation of entangled photon pairs, ensuring both energy and phase conservation, but at the cost of limited rate and degraded signal-to-noise ratio compared to laser-based interferometry approaches. We present a "quantum-like" nonlinear optical method that reaches super-resolution in single-photon detection regime. This is achieved by replacing photon-pairs by coherent states of light, mimicking quantum properties through classical nonlinear optics processes. Our scheme utilizes two high-brightness lasers. This results in a substantially greater signal-to-noise ratio compared to its quantum counterpart. Such an approach paves the way to significantly reduced acquisition times, providing a pathway to explore signals across a broader range of bandwidth. The need to increase the frequency bandwidth of the quantum sensor significantly motivates the potential applications of this pathway.

 \end{abstract}

\maketitle

\section{Introduction}
In the ever-evolving field of photonic quantum metrology, the ingenious harnessing of quantum properties of light paves the way for significative advances in the precise estimation of physical parameters~\cite{polino_photonic_2020}. This successful combination between photonics and quantum metrology permits pushing the boundaries of precision measurements forward, providing innovative avenues for scientific research and cutting-edge technological applications such as ghost-sensing~\cite{shapiro_physics_2012, padgett_introduction_2017}, optical quantum coherent tomography~\cite{teich_variations_2012, abouraddy_quantum-optical_2002, yepiz-graciano_quantum_2022}, spectroscopy~\cite{Mukamel_2020}, linear~\cite{LIGO} and nonlinear interferometry~\cite{chekhova_nonlinear_2016}. Nonlinear interferometers in quantum optics have received particular attention over the last few years. A striking property lies in induced coherence \cite{induced_coherence_PhysRevA.44.4614}, allowing the inference of optical phase at a wavelength different from the one detected. This enables for instance mid-infrared spectroscopy with near-infrared or visible light detection~\cite{PhysRevLett.67.318, kalashnikov_infrared_2016}. At low photon numbers, quantum optical typically arise as a direct consequence of parametric down-conversion processes occurring in a setup involving two nonlinear crystals. In such an architecture, a single incoming pump photon generates, in a delocalized manner across the two crystals, \textit{i.e.} without revealing the "which-source" information, a pair of photons—referred to as signal and idler photons—that exhibit strong correlations. Another advantage of these correlations is that single photons carries the signature of super-resolution, as usually measured in two-fold coincidence setups. This is the case since the creation probability associated with the generation is directly linked to the total phase of the interacting fields. In the case of spontaneous parametric down-conversion (SPDC), those fields are pump, signal and idler. Strikingly, mapping a two-photon phase onto that of single photons permits to reduce the experimental overhead, notably at the detection stage. In other words, replacing single-photon detectors and coincidence electronics by standard photodiodes enables both higher signal-to-noise ratios (SNR) and acquisition bandwidth~\cite{PhysRevLett.67.318, PhysRevLett.72.629, ono_observation_2019, vergyris_two-photon_2020}.\\
However, despite the quantum origin of these demonstrations, it was realized that many of the properties enabled by quantum interference could be demonstrated using classical correlations, freeing themselves from non-classical light generators and detectors~\cite{shapiro_classical_2015, wolf_spooky_2016, PhysRevLett.89.113601, PhysRevLett.93.093602, ko_emulating_2023, PhysRevA.74.041601, LeGouet:10, PhysRevLett.102.243601}.\\
Here, we introduce a nonlinear interferometry setup that demonstrates super-resolution in the single-photon detection regime, enabling measurement of optical properties at a non-detected wavelength. This phenomenon is demonstrated through a quantum-mimetic adaptation of the quantum approach, achieved by using simple lasers, frequency-engineered, to illuminate two nonlinear crystals arranged in a Mach-Zenhder interferometer. This emulation, achieved through sum-frequency generation, converts two pump photons in the near-IR into a single photon in the visible range, subsequently detected using a standard photodiode. We illustrate this concept through a typical example of optical phase estimation. A benchmarking experiment, dedicated to chromatic dispersion (CD) measurement, is performed and analyzed. Such estimation is essential in both classical communication and nonlinear optics at 1560~nm~\cite{agrawal_nonlinear_2013}. There, the CD is inferred from a measurement performed at 780~nm.

\section{Theoretical framework}

Super-resolution can be explained via the definition of maximally entangled state, that can be written as~\cite{noon_state_PhysRevLett.85.2733}:

\begin{equation}\label{eq: NOON state}
|\psi\rangle = \frac{|N,0\rangle_{ab} + e^{iN\phi}|0,N\rangle_{ab}}{\sqrt{2}}\, ,
\end{equation}

\begin{figure}[h!]
    \centering
    \includegraphics[width=1\linewidth]{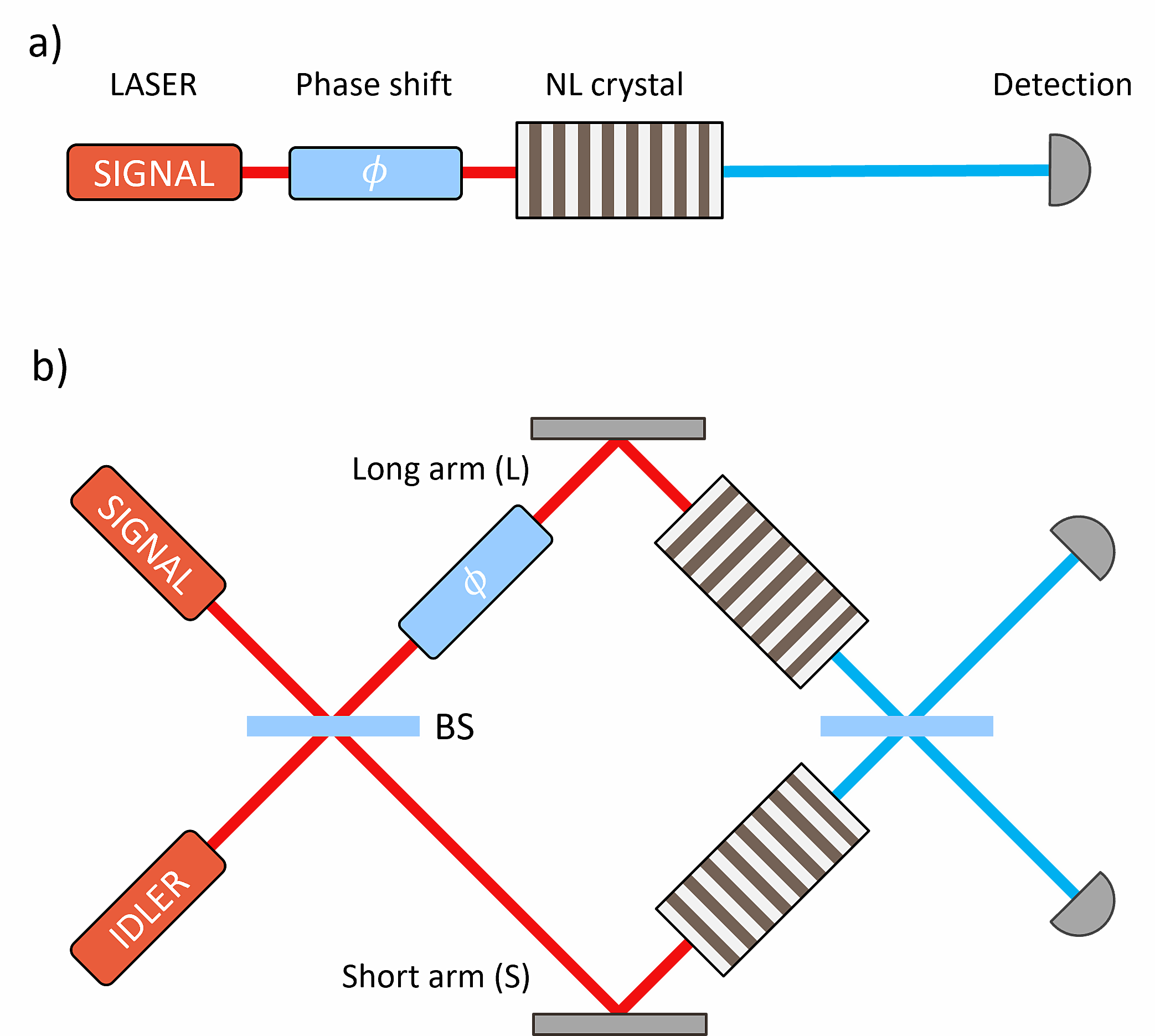}
    \caption{Schematic layout for super-resolved single-photon interferometry. a. A coherent state is directly sent to an NL crystal where stimulated parametric conversion merges N photons of the same energy into a single photon of higher energy. b. A coherent state is sent to a Mach-Zehnder interferometer where a NL medium is placed in each arms of the interferometer. If both arms are indistinguishable (same NL process with same efficiency), single-photon interferometry from converted light occurs at the second beam-splitter. Here, the relative phase acquired is equal to N times the phase acquired by the input state in the case of N${}^{th}$ harmonic generation, or equal to the sum of the phase acquired by both lasers in the case of sum frequency generation.}
    \label{fig: theoric setup}
\end{figure}

\noindent This state is referred to a N00N state, where $N$ and $\phi$ represent the number of photons and the relative phase between the two superposed states, respectively.  $a, b$ represent two photonic modes of an arbitrary orthogonal basis associated with polarization, path, or time-bin observable~\cite{noon_time_PhysRevLett.62.2205,noon_polar_PhysRevA.73.012316}. As a consequence of such a state-structure, the phase accumulated by the N-photon entangled state is multiplied by a factor N, a phenomenon referred to as super-resolution, widely used in quantum metrology \cite{metro_1_Unternahrer:18,metro_2_PhysRevResearch.3.033250,metro_3_Defienne2022}. In this particular case, the phase acquired by each photon is collective, as strong correlations bind them, as opposed to a collection of uncorrelated photons where each photon independently carries the acquired relative phase.\\
\figurename \ref{fig: theoric setup}.a is now considered for understanding super-resolution in single-photon interferometry. A coherent monochromatic photon source (called signal) of arbitrary frequency $\omega_s$  is defined by the state:

\begin{equation}\label{eq: coherent state}
    |\alpha_s, \omega_s\rangle = e^{\frac{-|\alpha_s|^2}{2}}\sum_{n=0}^\infty\frac{\alpha_s^n}{\sqrt{n!}}|n,\omega_s\rangle \, ,
\end{equation}
where $|\alpha_s|^2$ and $|n\rangle$ stand as the mean number of photon and the related states of the Fock basis, respectively. First, the coherent state is directed into a nonlinear (NL) crystal where stimulated parametric conversion takes place. By stimulated parametric conversion, we refer to a nonlinear optical process in which N photons interact simultaneously with the NL crystal and merge into a single photon of higher energy. The conservation of energy, expressed as $ N\omega_s  = \omega_p$ and phase, as $N\overrightarrow{k_s}  = \overrightarrow{k_p}$, relate to the N${}^{th}$ order harmonic generation. Thus, if the input state in the NL crystal is a coherent state, the converted output state will also be a coherent state, since the parametric conversion is a conservative process. By considering $\phi$ as the relative phase shift with respect to the initial coherent state, the conversion can be expressed as:

\begin{equation}\label{eq: NHG}
     \big|\lvert \alpha_s \rvert e^{i\phi}, \omega_s\rangle \Rightarrow  \big| \lvert \alpha_p \rvert e^{iN\phi}, \omega_p\rangle\, .
\end{equation}
We now consider the full setup shown in \figurename \ref{fig: theoric setup} b., with only the signal laser. The light is sent into an unbalanced Mach-Zehnder interferometer consisting of two balanced beam-splitters (BS). The top arm containing the relative phase difference is called long arm, and the bottom one is called short arm or reference arm.  Each of them contains a NL medium to ensure the parametric conversion. Within the interferometer, before arriving at the NL crystals, the state described by \equationautorefname{}~\eqref{eq: coherent state} becomes:

\begin{equation}\label{eq:4}
    |\psi\rangle = \frac{1}{\sqrt{2}}\bigl(|\alpha_s, \omega_s, S\rangle + e^{i\phi_s} |\alpha_s, \omega_s, L\rangle\bigr)\, ,
\end{equation}
where $S$ and $L$ correspond to the modes associated with the short and long arms, respectively (see \figurename{} \ref{fig: theoric setup}b.). Here, the phase term, $\phi_s$, is outside of the ket-vector as it corresponds to the relative phase between the superposed contributions. By combining the last equation with \equationautorefname{}~\eqref{eq: NHG}, the final state at the output of the interferometer reads:

\begin{equation}\label{eq: efficiencies}
    |\psi\rangle = \frac{1}{\sqrt{2}}\bigl(\sqrt{\eta}|\alpha_p,\omega_p,S\rangle + \sqrt{1-\eta}e^{iN\phi_s}|\alpha_p,\omega_p,L\rangle\bigr)\, ,
\end{equation}
where $\eta$ is the relative efficiency of the parametric conversion between the two arms. Finally, by considering the same efficiency in both arms and by discarding similar modes in the state (amplitude and frequency), we obtain:

\begin{equation}
    |\psi\rangle = \frac{1}{\sqrt{2}}\bigl(|S\rangle + e^{iN\phi_s}|L\rangle\bigr)\, ,
\end{equation}
which can be rewritten in a similar fashion as \equationautorefname{} \eqref{eq: NOON state}:

\begin{equation}\label{eq: NOON state single photon}
|\psi\rangle = \frac{|1,0\rangle_{SL} + e^{iN\phi_s}|0,1\rangle_{SL}}{\sqrt{2}}\, .
\end{equation}
The interference phenomenon pertains to single photons generated by stimulated parametric process in the NL crystal while the relative phase of the state is that of N entangled photons. In other words, the relative phase is the one of a N00N interference but super-resolution occurs via single-photon interference. This property derives from the fact that super-resolution is based on phase conservation of the parametric process, rather than quantum entanglement. In the case of N-photon interference, \textit{e.g.} two-photon interference from pairs generated by SPDC, super-resolution arises from phase conservation, while the induced correlations are carried by entanglement~\cite{noon_time_PhysRevLett.62.2205}.

Moreover, as the parametric process is stimulated (\textit{i.e.}, driven by a laser), the conversion efficiency is much greater than in the spontaneous regime, freeing the experiment for cumbersome detection system such as single photon detectors combined with time-to-digital converters to register coincidence events. \\
We emphasize that, in addition to being super-resolved, the interference pattern observed at the output of the interferometer provides information at the frequency of the photons used for the parametric conversion, rather than the frequency of the detected photons. This characteristic aligns with the principles of ghost sensing, where the properties of a sample under test are measured at one frequency while photon detection occurs at another~\cite{kalashnikov_infrared_2016}. However, it is essential to note a distinction between the two approaches. In our case, all photons actively contribute to the conversion process. In the context of ghost sensing, the photon that traverses the sample undergoes filtering subsequent to its passage through the interferometer, leveraging induced coherence from SU(1,1) type interferometers~\cite{su11_PhysRevA.33.4033}.\\

\begin{figure*}[ht!]
    \centering
    \includegraphics[width=0.8\linewidth]{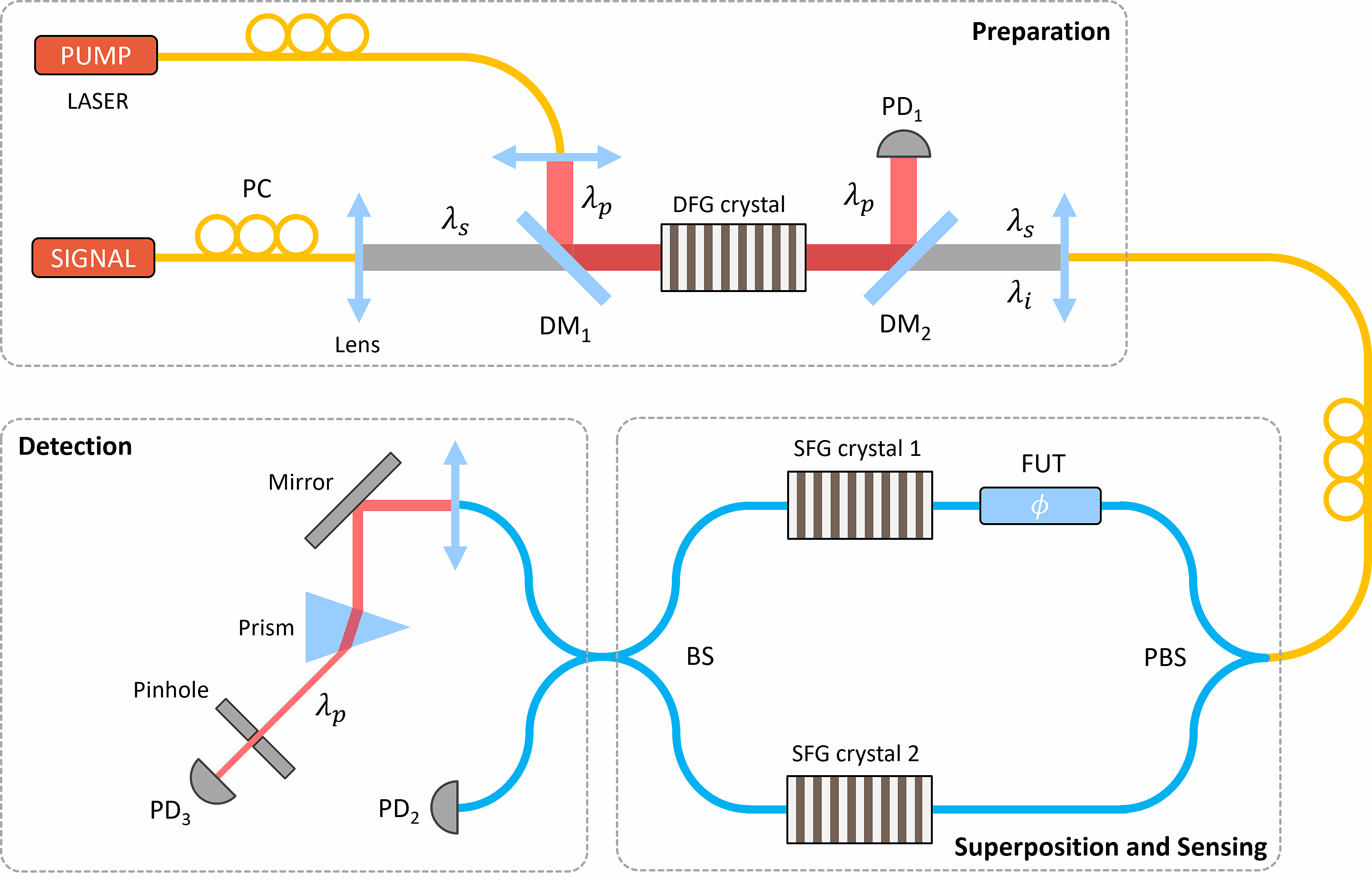}
    \caption{Experimental setup for the measurement of CD via super-resolved single-photon interferometry. 
    Preparation part: A frequency-locked Ti:SA laser operating in continuous-wave (CW) regime, with a pump wavelength of $\lambda_p = 780.3\,$nm, and a tunable C-band telecom laser ($\lambda_s$) are directed into a periodically poled Lithium Niobate (ppLN) crystal in a free-space configuration. DFG occurs in the crystal, generating a coherent in the telecom L-band ($\lambda_i$). To this end, the first dichroic mirror ($DM_1$) sends the visible light into the DFG crystal, while $DM_2$ ensures that no residual pump light passes through the rest of the setup. A photodiode ($PD_1$) is positioned after the second DM to monitor the power of the pump light. Superposition \& sensing part: The nonlinear (NL) interferometer is fully fibered and maintains polarization. A polarizing beam-splitter (PBS) operating in the telecom range and a beam-splitter (BS) operating in the visible range are placed at the entrance and exit of the interferometer, respectively. The fiber under test (FUT) is a commercial dispersion-compensated polarization-maintaining fiber. Fibered pigtailed ppLN waveguide crystals are placed in each arm of the interferometer. The related phase-matching conditions are designed to convert photons from the telecom C + L band back to the visible range $\lambda_{p}$ via SFG. Detection part: a free-space filtering stage consists of a prism and a pinhole to block any residual telecom light. An off-the-shelf photodiode ($PD_3$) records the interference pattern from the NL interferometer, while another photodiode ($PD_2$) enables monitoring visible light.}
    \label{fig: experimental setup}
\end{figure*}

As a proof-of-concept, we aim to demonstrate how our interferometric setup can be exploited for characterizing chromatic dispersion (group velocity dispersion, CD) at a telecom wavelength by measuring single-photon interference at a visible wavelength. We refer to a technique similar to quantum white light interferometry by exploiting entanglement to suppress odd-order dispersion terms, phenomenon called non-local dispersion cancellation~\cite{kaiser_quantum_2018, franson_disp_PhysRevA.45.3126}. Here, we show that this property can also be obtained by classical means and deduced from phase conservation in a single-photon interferometric experiment. To demonstrate this, a second laser, denoted idler, is introduced in the setup of \figurename \ref{fig: theoric setup}.b. The relative phase between the two arms of the interferometer is obtained from a dispersive medium. The initial state before the interferometer is defined as the tensor product of two coherent states at frequencies $\omega_s$ and $\omega_i$, where the subscript $i$ denotes the idler laser. Instead of considering $N^{th}$ order harmonic generation, focus is made on the sum-frequency generation process (SFG) that takes place in the two crystals. SFG is also a conservative process, governed by energy and phase conservation: $\omega_s + \omega_i  = \omega_p$ and $\overrightarrow{k_s} + \overrightarrow{k_i}  = \overrightarrow{k_p}$. Consequently, two coherent states undergoing SFG generate another coherent state. According to \equationautorefname{} \eqref{eq:4} and~\eqref{eq: efficiencies}, the output state of the interferometer can be expressed as:

\begin{equation}\label{eq: SFG state single photon}
|\psi\rangle = \frac{|1,0\rangle_{SL} + e^{i(\phi_s+\phi_i)}|0,1\rangle_{SL}}{\sqrt{2}}\, .
\end{equation}
Here, the relative phase carried by the SFG-generated single photons stands as the collective sum of the relative phase acquired by signal and ilder lasers within the interferometer. This term can be represented using a Taylor expansion centered around half the frequency of the converted light, say $\omega_0 = \frac{\omega_p}{2}$:

\begin{equation}\label{eq: phase SFG}
    \frac{\phi_s+\phi_i}{L}= \sum_{k=i,s}\sum_{n=0}^{\infty}\frac{\Delta\omega^n_k}{n!} \beta^{(n)} \, ,
\end{equation}
where $\beta^{(n)}=\left.\frac{\partial k}{\partial \omega}\right|_{\omega_0}$, $L$ and $\Delta\omega$ denote the length of the sample under test, the derivative of the wave vector, and the detuning from $\omega_0$, respectively. 
By imposing the frequency of the converted light to be constant while scanning both laser frequencies adequately, energy conservation ensures that $\Delta\omega_s = - \Delta\omega_i$. \equationautorefname{} \eqref{eq: phase SFG} therefore becomes:

\begin{equation}\label{eq:phasespectralQ}
    \frac{\phi_s + \phi_i}{L} = 2\beta^{(0)} + \beta^{(2)}\Delta\omega^2 +  o(\Delta\omega^4)\, .
\end{equation}

\noindent As can be seen, all odd-order dispersion terms vanish. The direct consequence of this cancellation lies in the transfer of non-local correlation, used in quantum white light interferometry, to classical (single photon) white light interferometry. This enables the direct extraction of CD from the measured interference pattern. Notably, phase conservation ensures that the phase of the converted light reflects the collective phase of both photons involved in the SFG process. Finally, the super-resolution property is highlighted by comparing \equationautorefname{} \eqref{eq:phasespectralQ} to its counterpart in a classical scheme~\cite{kaiser_quantum_2018}:

\begin{equation}\label{eq:phasespectralclass}
    \frac{\phi}{L} = \beta^{(0)} + \beta^{(1)} +  \frac{1}{2}  \beta^{(2)}\Delta\omega^2 +  \frac{1}{6}  \beta^{(3)}\Delta\omega^3 + o(\Delta\omega^4)\, .
\end{equation}

The absence of a factor $\frac{1}{2}$ in front of the term $\beta^{(2)}\Delta\omega^2$ in \equationautorefname{}~\eqref{eq:phasespectralQ} highlights the super-resolution effect for the quantum-like approach~\cite{kaiser_quantum_2018}.

\section{Experimental implementation}
As discussed above, a fundamental prerequisite lies in the energy conservation of signal and idler modes, which would be genuinely fulfilled  by a pair of entangled photons produced by SPDC. In the case of classical light, \textit{i.e.} with non-correlated photon pairs, this condition is not inherently satisfied. The most intuitive, however challenging, approach involves the use of two lasers that are perfectly frequency anti-correlated. An elegant strategy to overcome this constraint and to emulate conservation of energy is to rely on phase conservation in parametric conversion process. Based on difference frequency generation (DFG) occurring in a NL crystal, fed by two lasers —one wavelength-tuned and the other at a fixed wavelength— ensures that photons from the tunable laser and those generated by DFG genuinely fulfil the conservation law required for the experiment.\\

The experimental setup is shown in \figurename \ref{fig: experimental setup}. The upper part depicts the preparation of signal and idler modes, whose frequencies are anti-correlated. A frequency mode-locked Ti:Sa laser in CW regime in the visible range ($\lambda_p = 780.3\,$nm) and a tunable C-telecom band laser ($\lambda_s$) are combined in a NL crystal, labeled DFG crystal, to generate a coherent state frequency anti-correlated to the latter, belonging to the L-telecom band ($\lambda_i$). The visible laser plays the same role as the pump laser in an SPDC process granting energy conservation. Then, the two frequency-correlated coherent states are collected using an optical fiber towards the superposition \& sensing stage. This section consists of the NL interferometer, in which the device under test, here an optical fiber, is placed in one arm. The interferometer comprises one NL crystal, labeled SFG, in each of the two arms to generate photons at wavelength $\lambda_p$ in the superposed state defined by \equationautorefname{} \eqref{eq: SFG state single photon}. To ensure the same amount of converted light in both SFG crystals, a polarized beam-splitter and a polarization controller are positionned at the input of the interferometer. This permits to carefully adjust the polarization of the telecom light coming from the preparation part to compensate for any difference in the efficiency of the parametric processes between the two arms. The outputs of crystals are connected to a beam-splitter. The detection part allows the acquisition of an interference pattern by means of a standard silicon photodiode. Thus, the extracted CD value from the fiber under test (FUT) is given half the frequency of the pump laser while the measurement is made at $\lambda_p$. The photocurrent is analyzed with an oscilloscope triggered on the ramp of the tunable telecom laser to calibrate the measurement.\\

\begin{figure}[h]
    \centering
    \includegraphics[width=1\linewidth]{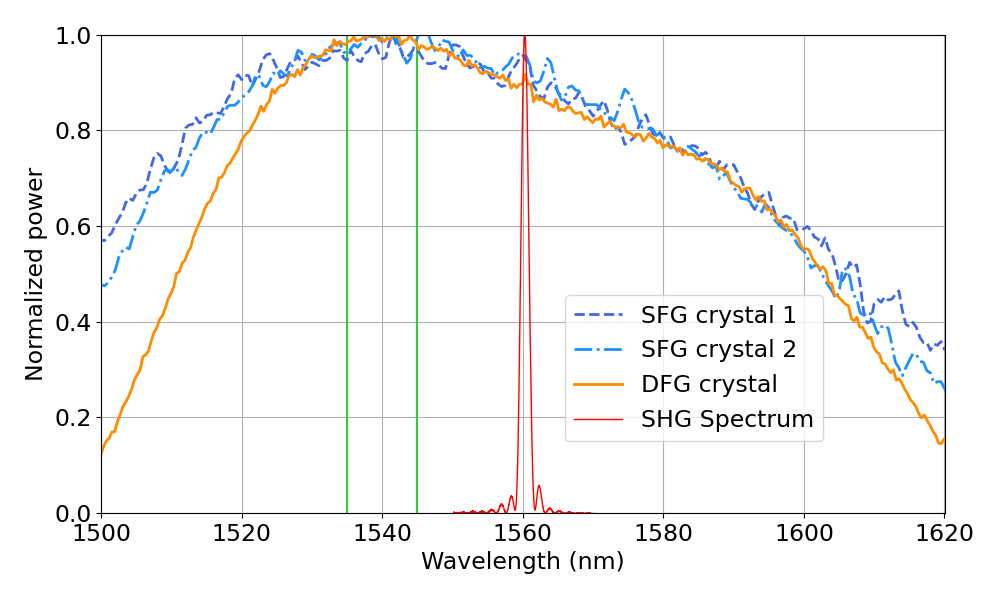}
    \caption{Measured  acceptances (spectra) of the crystals for different parametric processes. The blue curves represent the SFG acceptance from the crystals inside the interferometer. The orange curve stand for the DFG acceptance of the crystal needed for the generation of the idler coherent state. The red curve shows the SHG acceptance of one of the crystals inside the interferometer. The two vertical green lines show the spectral range of the measurement.}
    \label{fig: plage de fonctionnement}
\end{figure}

\noindent \figurename \ref{fig: plage de fonctionnement} illustrates the different wavelengths involved in the DFG and SFG processes. As these processes are cascaded, it is necessary to fulfill strict conditions regarding the wavelengths involved in these two distinct processes: (i) the telecom frequencies involved in the DFG process (orange curve) belong to within the spectrum of the two SFG processes (blue curves), and (ii) the pump wavelength must belong to the second harmonic generation (SHG) spectrum of the crystals placed inside the interferometer. Furthermore, attention must be paid to the telecom laser wavelength, which could lead to SHG by itself. In such a case, both degenerate SFG and SHG would mix, causing the interference pattern to vanish. To mitigate this unwanted parasitic SHG, the scanning range of the telecom laser, depicted in green in~\figurename \ref{fig: plage de fonctionnement}, has been selected to maximize the efficiency of the SFG process while keeping the efficiency of the SHG close to zero. Additionally, the filtering stage at the output of the interferometer enhances the suppression of potentially generated SHG parasitic light.\\

\noindent We highlight that the visibility of the interference pattern is strongly dependent to the overlap of the spectral distribution of the SFG process but remains independent of the DFG process. The contrast of the fringes is directly dependent on the efficiencies of the parametric processes within the interferometer, as shown in \equationautorefname{}~\eqref{eq: efficiencies}. Here, the spectral distributions of the SFG processes are nearly identical, ensuring a high-visibility pattern.

\begin{figure}[h]
    \centering
    \includegraphics[width=1\linewidth]{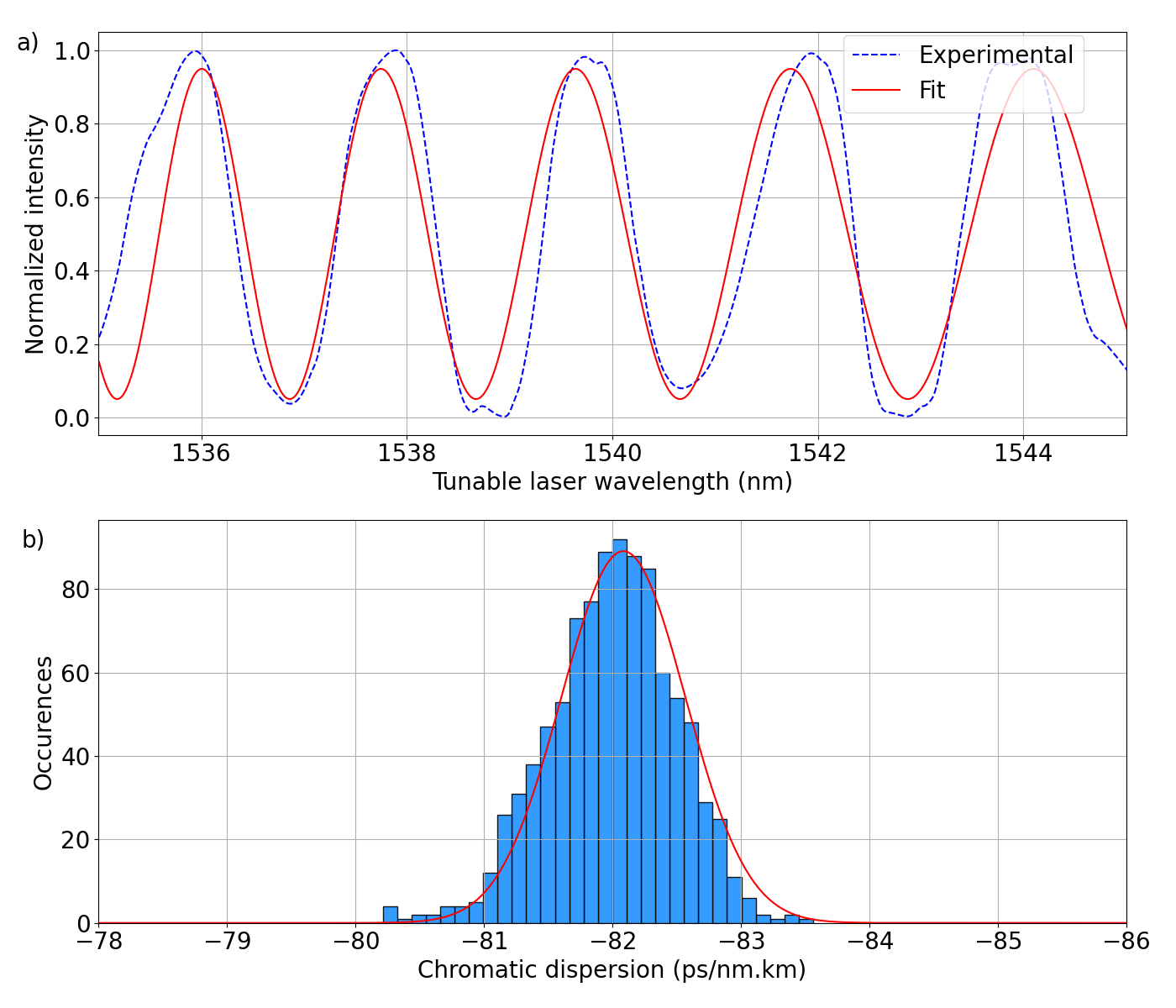}
    \caption{a) Interference pattern measured as a function of the tunable laser wavelength. The blue and red curves are the experimental values and a fit from \equationautorefname{} \eqref{eq: intensity} respectively. b) Histogram of the extracted CD for 1000 measurement. The red curve is the associated probability density function.}
    \label{fig:resultats}
\end{figure}

Hence, the intensity at the output of the system reads:

\begin{multline}\label{eq: intensity}
    I \propto sinc^2(\frac{\Delta k_{DFG} L_{DFG}}{2})sinc^2(\frac{\Delta k_{SFG} L_{SFG}}{2}) \\ \times\bigl[1+cos(\beta^{(2)}\Delta\omega^2 + 2\beta^{(0)})\bigr]\, ,
\end{multline}
    
\noindent where the first and second cardinal sines represent the phase matching of the DFG process and the SFG processes, respectively.
\figurename \ref{fig:resultats}.a displays the normalized interferogram acquired during a scan. The scanning speed of the C-telecom band tunable laser is $100\,$nm/s over the range $\lambda_s \in [1535;1545]\,$nm, resulting in a measurement time of $100\,$ms, significantly reducing the acquisition time compared to that required for a two-photon quantum white light interferometry experiment~\cite{kaiser_quantum_2018}.\\
As the scanning time is much shorter than the characteristic drift of the interferometer induced by environmental variations (mostly temperature), active stabilization is not required. Each dataset is fitted using the \equationautorefname{}~\eqref{eq: intensity} to extract the CD. As measurement methodology, we acquire 1000 scans and build a histogram of the CD values, depicted in \figurename \ref{fig:resultats}.b, in order to infer the statistical error. We obtain a value of $-82.08(1)\,\text{ps/(km.nm)}$, with a statistical error of $2.10^{-2}\%$, which lies within the manufacturer's specifications. This result is among the most precise reported to date in the literature, both in the classical~\cite{hlubina_chromatic_2013, jachura_high-visibility_2014, labonte_experimental_2006, grosz_measurement_2014, hlubina_differential_2007, galle_single-arm_2007, ye_dispersion_2002} and quantum regime~\cite{kaiser_quantum_2018}.

The precision of our approach is bounded by the shot-noise limit, which can be surpassed through the utilization of non-classical probes. Recent studies have harnessed the advantages of quantum nonlinear interferometry, leading to a significant enhancement of the measurement sensitivity beyond the shot-noise limit~\cite{qin_unconditional_2023, nold_quantum_2022}. However, these quantum resources are sensitive to losses~\cite{dorner_optimal_2009, demkowicz-dobrzanski_quantum_2009, kacprowicz_experimental_2010, datta_quantum_2011, escher_general_2011, thomas-peter_real-world_2011}. There are several scenarios in which our classical approach could offer advantages over the quantum case, particularly in domains where inherent losses are significant, as is the case in spectroscopy~\cite{Mukamel_2020} and imaging with undetected photons~\cite{lemos_quantum_2014} or absorption measurements~\cite{crespi_measuring_2012}. Moreover, the capability to discern optical phase at a wavelength different from the one being measured is particularly valuable in these fields, notably in the mid-IR~\cite{kalashnikov_infrared_2016}.\\
We have chosen to measure CD through a spectral method as a proof-of-principle, requiring a static measurement where the phase term within the interferometer remains constant over time. However, let us emphasize our capability to perform dynamic measurements, which represents a current challenge for quantum sensors relying on entangled photon pairs due to their moderate coincidence counting rates, limiting signal acquisition to the kHz range~\cite{chen_quantum_2022}. In our case, detecting photons using a photodiode removes this constraint, allowing us to exploit the photodiode's bandwidth, which can extend up to several 10s of GHz.\\
Finally, the performance of the experiment could be easily augmented by harnessing the second output of the NL interferometer, adding a filtering stage and a photodiode. Conventional homodyne detection can be implemented, enabling the suppression of common noise, such as phase and amplitude noise from the laser, and directly acquiring the normalization of the two-photon interference pattern~\cite{nold_quantum_2022}.

\section{Conclusion}

This work has introduced a NL interferometer illuminated by classical light, allowing super-resolved measurement inherent to two-photon interferometry in single-photon applications. By exploiting phase and energy conservation in NL parametric processes, we successfully demonstrate the extraction of chromatic dispersion, as a proof-of-concept of the method, through a dispersion cancellation phenomenon. The efficiency of the stimulated processes, coupled to classical detection methods, enables fast measurement schemes, paving the way for dynamic measurements. In a broader context, this NL interferometry method, based on degenerate SFG, promises applications beyond CD measurements, potentially extending to ghost sensing detection using other NL processes like DFG or non-degenerated SFG, without the need for induced coherence. The versatility and efficiency of this quantum-like approach offer practical advantages for real-world applications, providing an alternative and complementary way to sensors based on non-classical light in diverse fields such as spectroscopy and imaging.

\section*{Acknowledgment}
This work has been conducted within the framework of the project OPTIMAL granted by the European Union by means of the Fond Européen de développement régional (FEDER). The authors also acknowledge financial support from the Agence Nationale de la Recherche (ANR) through the projects METROPOLIS (ANR-19-CE47-0008), QAFEINE (21-ASTR-0007-DA), PARADIS (ANR-22-ASTR-0027-01), ADEQUADE (European Defense Fund, 2023), and the French government through its Investments for the Future programme under the Université Côte d'Azur UCA-JEDI project (Quantum@UCA) managed by the ANR (ANR-15-IDEX-01).\\

\section*{Author information}
All the authors contributed equally to the entire process, from the first draft to the final version of the manuscript before submission. We all read, discussed, and contributed to the writing, reviewing, and editing. LL \& ST  coordinated and managed the project, ensuring its successful completion.

\section*{Competing interests}
The authors declare no competing interests.

\section*{Data availability}
The datasets used and analysed during the current study available from the corresponding author on reasonable request.


\begin{thebibliography}{51}%
\makeatletter
\providecommand \@ifxundefined [1]{%
 \@ifx{#1\undefined}
}%
\providecommand \@ifnum [1]{%
 \ifnum #1\expandafter \@firstoftwo
 \else \expandafter \@secondoftwo
 \fi
}%
\providecommand \@ifx [1]{%
 \ifx #1\expandafter \@firstoftwo
 \else \expandafter \@secondoftwo
 \fi
}%
\providecommand \natexlab [1]{#1}%
\providecommand \enquote  [1]{``#1''}%
\providecommand \bibnamefont  [1]{#1}%
\providecommand \bibfnamefont [1]{#1}%
\providecommand \citenamefont [1]{#1}%
\providecommand \href@noop [0]{\@secondoftwo}%
\providecommand \href [0]{\begingroup \@sanitize@url \@href}%
\providecommand \@href[1]{\@@startlink{#1}\@@href}%
\providecommand \@@href[1]{\endgroup#1\@@endlink}%
\providecommand \@sanitize@url [0]{\catcode `\\12\catcode `\$12\catcode `\&12\catcode `\#12\catcode `\^12\catcode `\_12\catcode `\%12\relax}%
\providecommand \@@startlink[1]{}%
\providecommand \@@endlink[0]{}%
\providecommand \url  [0]{\begingroup\@sanitize@url \@url }%
\providecommand \@url [1]{\endgroup\@href {#1}{\urlprefix }}%
\providecommand \urlprefix  [0]{URL }%
\providecommand \Eprint [0]{\href }%
\providecommand \doibase [0]{http://dx.doi.org/}%
\providecommand \selectlanguage [0]{\@gobble}%
\providecommand \bibinfo  [0]{\@secondoftwo}%
\providecommand \bibfield  [0]{\@secondoftwo}%
\providecommand \translation [1]{[#1]}%
\providecommand \BibitemOpen [0]{}%
\providecommand \bibitemStop [0]{}%
\providecommand \bibitemNoStop [0]{.\EOS\space}%
\providecommand \EOS [0]{\spacefactor3000\relax}%
\providecommand \BibitemShut  [1]{\csname bibitem#1\endcsname}%
\let\auto@bib@innerbib\@empty
\bibitem [{\citenamefont {Polino}\ \emph {et~al.}(2020)\citenamefont {Polino}, \citenamefont {Valeri}, \citenamefont {Spagnolo},\ and\ \citenamefont {Sciarrino}}]{polino_photonic_2020}%
  \BibitemOpen
  \bibfield  {author} {\bibinfo {author} {\bibfnamefont {E.}~\bibnamefont {Polino}}, \bibinfo {author} {\bibfnamefont {M.}~\bibnamefont {Valeri}}, \bibinfo {author} {\bibfnamefont {N.}~\bibnamefont {Spagnolo}}, \ and\ \bibinfo {author} {\bibfnamefont {F.}~\bibnamefont {Sciarrino}},\ }\href {\doibase 10.1116/5.0007577} {\bibfield  {journal} {\bibinfo  {journal} {AVS Quantum Science}\ }\textbf {\bibinfo {volume} {2}},\ \bibinfo {pages} {024703} (\bibinfo {year} {2020})}\BibitemShut {NoStop}%
\bibitem [{\citenamefont {Shapiro}\ and\ \citenamefont {Boyd}(2012)}]{shapiro_physics_2012}%
  \BibitemOpen
  \bibfield  {author} {\bibinfo {author} {\bibfnamefont {J.~H.}\ \bibnamefont {Shapiro}}\ and\ \bibinfo {author} {\bibfnamefont {R.~W.}\ \bibnamefont {Boyd}},\ }\href {\doibase 10.1007/s11128-011-0356-5} {\bibfield  {journal} {\bibinfo  {journal} {Quantum Inf Process}\ }\textbf {\bibinfo {volume} {11}},\ \bibinfo {pages} {949} (\bibinfo {year} {2012})}\BibitemShut {NoStop}%
\bibitem [{\citenamefont {Padgett}\ and\ \citenamefont {Boyd}(2017)}]{padgett_introduction_2017}%
  \BibitemOpen
  \bibfield  {author} {\bibinfo {author} {\bibfnamefont {M.~J.}\ \bibnamefont {Padgett}}\ and\ \bibinfo {author} {\bibfnamefont {R.~W.}\ \bibnamefont {Boyd}},\ }\href {\doibase 10.1098/rsta.2016.0233} {\bibfield  {journal} {\bibinfo  {journal} {Philos Trans A Math Phys Eng Sci}\ }\textbf {\bibinfo {volume} {375}} (\bibinfo {year} {2017}),\ 10.1098/rsta.2016.0233}\BibitemShut {NoStop}%
\bibitem [{\citenamefont {Teich}\ \emph {et~al.}(2012)\citenamefont {Teich}, \citenamefont {Saleh}, \citenamefont {Wong},\ and\ \citenamefont {Shapiro}}]{teich_variations_2012}%
  \BibitemOpen
  \bibfield  {author} {\bibinfo {author} {\bibfnamefont {M.~C.}\ \bibnamefont {Teich}}, \bibinfo {author} {\bibfnamefont {B.~E.~A.}\ \bibnamefont {Saleh}}, \bibinfo {author} {\bibfnamefont {F.~N.~C.}\ \bibnamefont {Wong}}, \ and\ \bibinfo {author} {\bibfnamefont {J.~H.}\ \bibnamefont {Shapiro}},\ }\href {\doibase 10.1007/s11128-011-0266-6} {\bibfield  {journal} {\bibinfo  {journal} {Quantum Inf Process}\ }\textbf {\bibinfo {volume} {11}},\ \bibinfo {pages} {903} (\bibinfo {year} {2012})}\BibitemShut {NoStop}%
\bibitem [{\citenamefont {Abouraddy}\ \emph {et~al.}(2002)\citenamefont {Abouraddy}, \citenamefont {Nasr}, \citenamefont {Saleh}, \citenamefont {Sergienko},\ and\ \citenamefont {Teich}}]{abouraddy_quantum-optical_2002}%
  \BibitemOpen
  \bibfield  {author} {\bibinfo {author} {\bibfnamefont {A.~F.}\ \bibnamefont {Abouraddy}}, \bibinfo {author} {\bibfnamefont {M.~B.}\ \bibnamefont {Nasr}}, \bibinfo {author} {\bibfnamefont {B.~E.~A.}\ \bibnamefont {Saleh}}, \bibinfo {author} {\bibfnamefont {A.~V.}\ \bibnamefont {Sergienko}}, \ and\ \bibinfo {author} {\bibfnamefont {M.~C.}\ \bibnamefont {Teich}},\ }\href {\doibase 10.1103/PhysRevA.65.053817} {\bibfield  {journal} {\bibinfo  {journal} {Physical Review A}\ }\textbf {\bibinfo {volume} {65}} (\bibinfo {year} {2002}),\ 10.1103/PhysRevA.65.053817}\BibitemShut {NoStop}%
\bibitem [{\citenamefont {Yepiz-Graciano}\ \emph {et~al.}(2022)\citenamefont {Yepiz-Graciano}, \citenamefont {Ibarra-Borja}, \citenamefont {Ramírez~Alarcón}, \citenamefont {Gutiérrez-Torres}, \citenamefont {Cruz-Ramírez}, \citenamefont {Lopez-Mago},\ and\ \citenamefont {U’Ren}}]{yepiz-graciano_quantum_2022}%
  \BibitemOpen
  \bibfield  {author} {\bibinfo {author} {\bibfnamefont {P.}~\bibnamefont {Yepiz-Graciano}}, \bibinfo {author} {\bibfnamefont {Z.}~\bibnamefont {Ibarra-Borja}}, \bibinfo {author} {\bibfnamefont {R.}~\bibnamefont {Ramírez~Alarcón}}, \bibinfo {author} {\bibfnamefont {G.}~\bibnamefont {Gutiérrez-Torres}}, \bibinfo {author} {\bibfnamefont {H.}~\bibnamefont {Cruz-Ramírez}}, \bibinfo {author} {\bibfnamefont {D.}~\bibnamefont {Lopez-Mago}}, \ and\ \bibinfo {author} {\bibfnamefont {A.~B.}\ \bibnamefont {U’Ren}},\ }\href {\doibase 10.1103/PhysRevApplied.18.034060} {\bibfield  {journal} {\bibinfo  {journal} {Phys. Rev. Appl.}\ }\textbf {\bibinfo {volume} {18}},\ \bibinfo {pages} {034060} (\bibinfo {year} {2022})},\ \bibinfo {note} {publisher: American Physical Society}\BibitemShut {NoStop}%
\bibitem [{\citenamefont {Mukamel}\ \emph {et~al.}(2020)\citenamefont {Mukamel}, \citenamefont {Freyberger}, \citenamefont {Schleich}, \citenamefont {Bellini}, \citenamefont {Zavatta}, \citenamefont {Leuchs}, \citenamefont {Silberhorn}, \citenamefont {Boyd}, \citenamefont {Sánchez-Soto}, \citenamefont {Stefanov}, \citenamefont {Barbieri}, \citenamefont {Paterova}, \citenamefont {Krivitsky}, \citenamefont {Shwartz}, \citenamefont {Tamasaku}, \citenamefont {Dorfman}, \citenamefont {Schlawin}, \citenamefont {Sandoghdar}, \citenamefont {Raymer}, \citenamefont {Marcus}, \citenamefont {Varnavski}, \citenamefont {Goodson}, \citenamefont {Zhou}, \citenamefont {Shi}, \citenamefont {Asban}, \citenamefont {Scully}, \citenamefont {Agarwal}, \citenamefont {Peng}, \citenamefont {Sokolov}, \citenamefont {Zhang}, \citenamefont {Zubairy}, \citenamefont {Vartanyants}, \citenamefont {del Valle},\ and\ \citenamefont {Laussy}}]{Mukamel_2020}%
  \BibitemOpen
  \bibfield  {author} {\bibinfo {author} {\bibfnamefont {S.}~\bibnamefont {Mukamel}}, \bibinfo {author} {\bibfnamefont {M.}~\bibnamefont {Freyberger}}, \bibinfo {author} {\bibfnamefont {W.}~\bibnamefont {Schleich}}, \bibinfo {author} {\bibfnamefont {M.}~\bibnamefont {Bellini}}, \bibinfo {author} {\bibfnamefont {A.}~\bibnamefont {Zavatta}}, \bibinfo {author} {\bibfnamefont {G.}~\bibnamefont {Leuchs}}, \bibinfo {author} {\bibfnamefont {C.}~\bibnamefont {Silberhorn}}, \bibinfo {author} {\bibfnamefont {R.~W.}\ \bibnamefont {Boyd}}, \bibinfo {author} {\bibfnamefont {L.~L.}\ \bibnamefont {Sánchez-Soto}}, \bibinfo {author} {\bibfnamefont {A.}~\bibnamefont {Stefanov}}, \bibinfo {author} {\bibfnamefont {M.}~\bibnamefont {Barbieri}}, \bibinfo {author} {\bibfnamefont {A.}~\bibnamefont {Paterova}}, \bibinfo {author} {\bibfnamefont {L.}~\bibnamefont {Krivitsky}}, \bibinfo {author} {\bibfnamefont {S.}~\bibnamefont {Shwartz}}, \bibinfo {author} {\bibfnamefont {K.}~\bibnamefont {Tamasaku}}, \bibinfo {author} {\bibfnamefont
  {K.}~\bibnamefont {Dorfman}}, \bibinfo {author} {\bibfnamefont {F.}~\bibnamefont {Schlawin}}, \bibinfo {author} {\bibfnamefont {V.}~\bibnamefont {Sandoghdar}}, \bibinfo {author} {\bibfnamefont {M.}~\bibnamefont {Raymer}}, \bibinfo {author} {\bibfnamefont {A.}~\bibnamefont {Marcus}}, \bibinfo {author} {\bibfnamefont {O.}~\bibnamefont {Varnavski}}, \bibinfo {author} {\bibfnamefont {T.}~\bibnamefont {Goodson}}, \bibinfo {author} {\bibfnamefont {Z.-Y.}\ \bibnamefont {Zhou}}, \bibinfo {author} {\bibfnamefont {B.-S.}\ \bibnamefont {Shi}}, \bibinfo {author} {\bibfnamefont {S.}~\bibnamefont {Asban}}, \bibinfo {author} {\bibfnamefont {M.}~\bibnamefont {Scully}}, \bibinfo {author} {\bibfnamefont {G.}~\bibnamefont {Agarwal}}, \bibinfo {author} {\bibfnamefont {T.}~\bibnamefont {Peng}}, \bibinfo {author} {\bibfnamefont {A.~V.}\ \bibnamefont {Sokolov}}, \bibinfo {author} {\bibfnamefont {Z.-D.}\ \bibnamefont {Zhang}}, \bibinfo {author} {\bibfnamefont {M.~S.}\ \bibnamefont {Zubairy}}, \bibinfo {author} {\bibfnamefont
  {I.~A.}\ \bibnamefont {Vartanyants}}, \bibinfo {author} {\bibfnamefont {E.}~\bibnamefont {del Valle}}, \ and\ \bibinfo {author} {\bibfnamefont {F.}~\bibnamefont {Laussy}},\ }\href {\doibase 10.1088/1361-6455/ab69a8} {\bibfield  {journal} {\bibinfo  {journal} {Journal of Physics B: Atomic, Molecular and Optical Physics}\ }\textbf {\bibinfo {volume} {53}},\ \bibinfo {pages} {072002} (\bibinfo {year} {2020})}\BibitemShut {NoStop}%
\bibitem [{\citenamefont {et~al.}(2023)}]{LIGO}%
  \BibitemOpen
  \bibfield  {author} {\bibinfo {author} {\bibfnamefont {G.}~\bibnamefont {et~al.}} (\bibinfo {collaboration} {LIGO O4 Detector Collaboration}),\ }\href {\doibase 10.1103/PhysRevX.13.041021} {\bibfield  {journal} {\bibinfo  {journal} {Phys. Rev. X}\ }\textbf {\bibinfo {volume} {13}},\ \bibinfo {pages} {041021} (\bibinfo {year} {2023})}\BibitemShut {NoStop}%
\bibitem [{\citenamefont {Chekhova}\ and\ \citenamefont {Ou}(2016)}]{chekhova_nonlinear_2016}%
  \BibitemOpen
  \bibfield  {author} {\bibinfo {author} {\bibfnamefont {M.~V.}\ \bibnamefont {Chekhova}}\ and\ \bibinfo {author} {\bibfnamefont {Z.~Y.}\ \bibnamefont {Ou}},\ }\href {\doibase 10.1364/AOP.8.000104} {\bibfield  {journal} {\bibinfo  {journal} {Adv. Opt. Photon.}\ }\textbf {\bibinfo {volume} {8}},\ \bibinfo {pages} {104} (\bibinfo {year} {2016})}\BibitemShut {NoStop}%
\bibitem [{\citenamefont {Wang}\ \emph {et~al.}(1991)\citenamefont {Wang}, \citenamefont {Zou},\ and\ \citenamefont {Mandel}}]{induced_coherence_PhysRevA.44.4614}%
  \BibitemOpen
  \bibfield  {author} {\bibinfo {author} {\bibfnamefont {L.~J.}\ \bibnamefont {Wang}}, \bibinfo {author} {\bibfnamefont {X.~Y.}\ \bibnamefont {Zou}}, \ and\ \bibinfo {author} {\bibfnamefont {L.}~\bibnamefont {Mandel}},\ }\href {\doibase 10.1103/PhysRevA.44.4614} {\bibfield  {journal} {\bibinfo  {journal} {Phys. Rev. A}\ }\textbf {\bibinfo {volume} {44}},\ \bibinfo {pages} {4614} (\bibinfo {year} {1991})}\BibitemShut {NoStop}%
\bibitem [{\citenamefont {Zou}\ \emph {et~al.}(1991)\citenamefont {Zou}, \citenamefont {Wang},\ and\ \citenamefont {Mandel}}]{PhysRevLett.67.318}%
  \BibitemOpen
  \bibfield  {author} {\bibinfo {author} {\bibfnamefont {X.~Y.}\ \bibnamefont {Zou}}, \bibinfo {author} {\bibfnamefont {L.~J.}\ \bibnamefont {Wang}}, \ and\ \bibinfo {author} {\bibfnamefont {L.}~\bibnamefont {Mandel}},\ }\href {\doibase 10.1103/PhysRevLett.67.318} {\bibfield  {journal} {\bibinfo  {journal} {Phys. Rev. Lett.}\ }\textbf {\bibinfo {volume} {67}},\ \bibinfo {pages} {318} (\bibinfo {year} {1991})}\BibitemShut {NoStop}%
\bibitem [{\citenamefont {Kalashnikov}\ \emph {et~al.}(2016)\citenamefont {Kalashnikov}, \citenamefont {Paterova}, \citenamefont {Kulik},\ and\ \citenamefont {Krivitsky}}]{kalashnikov_infrared_2016}%
  \BibitemOpen
  \bibfield  {author} {\bibinfo {author} {\bibfnamefont {D.~A.}\ \bibnamefont {Kalashnikov}}, \bibinfo {author} {\bibfnamefont {A.~V.}\ \bibnamefont {Paterova}}, \bibinfo {author} {\bibfnamefont {S.~P.}\ \bibnamefont {Kulik}}, \ and\ \bibinfo {author} {\bibfnamefont {L.~A.}\ \bibnamefont {Krivitsky}},\ }\href {\doibase 10.1038/nphoton.2015.252} {\bibfield  {journal} {\bibinfo  {journal} {Nature Photonics}\ }\textbf {\bibinfo {volume} {10}},\ \bibinfo {pages} {98} (\bibinfo {year} {2016})},\ \bibinfo {note} {arXiv: 1506.07223}\BibitemShut {NoStop}%
\bibitem [{\citenamefont {Herzog}\ \emph {et~al.}(1994)\citenamefont {Herzog}, \citenamefont {Rarity}, \citenamefont {Weinfurter},\ and\ \citenamefont {Zeilinger}}]{PhysRevLett.72.629}%
  \BibitemOpen
  \bibfield  {author} {\bibinfo {author} {\bibfnamefont {T.~J.}\ \bibnamefont {Herzog}}, \bibinfo {author} {\bibfnamefont {J.~G.}\ \bibnamefont {Rarity}}, \bibinfo {author} {\bibfnamefont {H.}~\bibnamefont {Weinfurter}}, \ and\ \bibinfo {author} {\bibfnamefont {A.}~\bibnamefont {Zeilinger}},\ }\href {\doibase 10.1103/PhysRevLett.72.629} {\bibfield  {journal} {\bibinfo  {journal} {Phys. Rev. Lett.}\ }\textbf {\bibinfo {volume} {72}},\ \bibinfo {pages} {629} (\bibinfo {year} {1994})}\BibitemShut {NoStop}%
\bibitem [{\citenamefont {Ono}\ \emph {et~al.}(2019)\citenamefont {Ono}, \citenamefont {Sinclair}, \citenamefont {Bonneau}, \citenamefont {Thompson}, \citenamefont {Matthews},\ and\ \citenamefont {Rarity}}]{ono_observation_2019}%
  \BibitemOpen
  \bibfield  {author} {\bibinfo {author} {\bibfnamefont {T.}~\bibnamefont {Ono}}, \bibinfo {author} {\bibfnamefont {G.~F.}\ \bibnamefont {Sinclair}}, \bibinfo {author} {\bibfnamefont {D.}~\bibnamefont {Bonneau}}, \bibinfo {author} {\bibfnamefont {M.~G.}\ \bibnamefont {Thompson}}, \bibinfo {author} {\bibfnamefont {J.~C.~F.}\ \bibnamefont {Matthews}}, \ and\ \bibinfo {author} {\bibfnamefont {J.~G.}\ \bibnamefont {Rarity}},\ }\href {\doibase 10.1364/OL.44.001277} {\bibfield  {journal} {\bibinfo  {journal} {Opt. Lett., OL}\ }\textbf {\bibinfo {volume} {44}},\ \bibinfo {pages} {1277} (\bibinfo {year} {2019})},\ \bibinfo {note} {publisher: Optical Society of America}\BibitemShut {NoStop}%
\bibitem [{\citenamefont {Vergyris}\ \emph {et~al.}(2020)\citenamefont {Vergyris}, \citenamefont {Babin}, \citenamefont {Nold}, \citenamefont {Gouzien}, \citenamefont {Herrmann}, \citenamefont {Silberhorn}, \citenamefont {Alibart}, \citenamefont {Tanzilli},\ and\ \citenamefont {Kaiser}}]{vergyris_two-photon_2020}%
  \BibitemOpen
  \bibfield  {author} {\bibinfo {author} {\bibfnamefont {P.}~\bibnamefont {Vergyris}}, \bibinfo {author} {\bibfnamefont {C.}~\bibnamefont {Babin}}, \bibinfo {author} {\bibfnamefont {R.}~\bibnamefont {Nold}}, \bibinfo {author} {\bibfnamefont {E.}~\bibnamefont {Gouzien}}, \bibinfo {author} {\bibfnamefont {H.}~\bibnamefont {Herrmann}}, \bibinfo {author} {\bibfnamefont {C.}~\bibnamefont {Silberhorn}}, \bibinfo {author} {\bibfnamefont {O.}~\bibnamefont {Alibart}}, \bibinfo {author} {\bibfnamefont {S.}~\bibnamefont {Tanzilli}}, \ and\ \bibinfo {author} {\bibfnamefont {F.}~\bibnamefont {Kaiser}},\ }\href {\doibase 10.1063/5.0009527} {\bibfield  {journal} {\bibinfo  {journal} {Appl. Phys. Lett.}\ }\textbf {\bibinfo {volume} {117}},\ \bibinfo {pages} {024001} (\bibinfo {year} {2020})},\ \bibinfo {note} {publisher: American Institute of Physics}\BibitemShut {NoStop}%
\bibitem [{\citenamefont {Shapiro}\ \emph {et~al.}(2015)\citenamefont {Shapiro}, \citenamefont {Venkatraman},\ and\ \citenamefont {Wong}}]{shapiro_classical_2015}%
  \BibitemOpen
  \bibfield  {author} {\bibinfo {author} {\bibfnamefont {J.~H.}\ \bibnamefont {Shapiro}}, \bibinfo {author} {\bibfnamefont {D.}~\bibnamefont {Venkatraman}}, \ and\ \bibinfo {author} {\bibfnamefont {F.~N.~C.}\ \bibnamefont {Wong}},\ }\href {\doibase 10.1038/srep10329} {\bibfield  {journal} {\bibinfo  {journal} {Sci Rep}\ }\textbf {\bibinfo {volume} {5}},\ \bibinfo {pages} {10329} (\bibinfo {year} {2015})}\BibitemShut {NoStop}%
\bibitem [{\citenamefont {Wolf}\ and\ \citenamefont {Silberberg}(2016)}]{wolf_spooky_2016}%
  \BibitemOpen
  \bibfield  {author} {\bibinfo {author} {\bibfnamefont {J.-P.}\ \bibnamefont {Wolf}}\ and\ \bibinfo {author} {\bibfnamefont {Y.}~\bibnamefont {Silberberg}},\ }\href {\doibase 10.1038/nphoton.2015.267} {\bibfield  {journal} {\bibinfo  {journal} {Nature Photon}\ }\textbf {\bibinfo {volume} {10}},\ \bibinfo {pages} {77} (\bibinfo {year} {2016})}\BibitemShut {NoStop}%
\bibitem [{\citenamefont {Bennink}\ \emph {et~al.}(2002)\citenamefont {Bennink}, \citenamefont {Bentley},\ and\ \citenamefont {Boyd}}]{PhysRevLett.89.113601}%
  \BibitemOpen
  \bibfield  {author} {\bibinfo {author} {\bibfnamefont {R.~S.}\ \bibnamefont {Bennink}}, \bibinfo {author} {\bibfnamefont {S.~J.}\ \bibnamefont {Bentley}}, \ and\ \bibinfo {author} {\bibfnamefont {R.~W.}\ \bibnamefont {Boyd}},\ }\href {\doibase 10.1103/PhysRevLett.89.113601} {\bibfield  {journal} {\bibinfo  {journal} {Phys. Rev. Lett.}\ }\textbf {\bibinfo {volume} {89}},\ \bibinfo {pages} {113601} (\bibinfo {year} {2002})}\BibitemShut {NoStop}%
\bibitem [{\citenamefont {Gatti}\ \emph {et~al.}(2004)\citenamefont {Gatti}, \citenamefont {Brambilla}, \citenamefont {Bache},\ and\ \citenamefont {Lugiato}}]{PhysRevLett.93.093602}%
  \BibitemOpen
  \bibfield  {author} {\bibinfo {author} {\bibfnamefont {A.}~\bibnamefont {Gatti}}, \bibinfo {author} {\bibfnamefont {E.}~\bibnamefont {Brambilla}}, \bibinfo {author} {\bibfnamefont {M.}~\bibnamefont {Bache}}, \ and\ \bibinfo {author} {\bibfnamefont {L.~A.}\ \bibnamefont {Lugiato}},\ }\href {\doibase 10.1103/PhysRevLett.93.093602} {\bibfield  {journal} {\bibinfo  {journal} {Phys. Rev. Lett.}\ }\textbf {\bibinfo {volume} {93}},\ \bibinfo {pages} {093602} (\bibinfo {year} {2004})}\BibitemShut {NoStop}%
\bibitem [{\citenamefont {Ko}\ \emph {et~al.}(2023)\citenamefont {Ko}, \citenamefont {Cook},\ and\ \citenamefont {Whaley}}]{ko_emulating_2023}%
  \BibitemOpen
  \bibfield  {author} {\bibinfo {author} {\bibfnamefont {L.}~\bibnamefont {Ko}}, \bibinfo {author} {\bibfnamefont {R.~L.}\ \bibnamefont {Cook}}, \ and\ \bibinfo {author} {\bibfnamefont {K.~B.}\ \bibnamefont {Whaley}},\ }\href {\doibase 10.1021/acs.jpclett.3c01714} {\bibfield  {journal} {\bibinfo  {journal} {J. Phys. Chem. Lett.}\ }\textbf {\bibinfo {volume} {14}},\ \bibinfo {pages} {8050} (\bibinfo {year} {2023})},\ \bibinfo {note} {publisher: American Chemical Society}\BibitemShut {NoStop}%
\bibitem [{\citenamefont {Erkmen}\ and\ \citenamefont {Shapiro}(2006)}]{PhysRevA.74.041601}%
  \BibitemOpen
  \bibfield  {author} {\bibinfo {author} {\bibfnamefont {B.~I.}\ \bibnamefont {Erkmen}}\ and\ \bibinfo {author} {\bibfnamefont {J.~H.}\ \bibnamefont {Shapiro}},\ }\href {\doibase 10.1103/PhysRevA.74.041601} {\bibfield  {journal} {\bibinfo  {journal} {Phys. Rev. A}\ }\textbf {\bibinfo {volume} {74}},\ \bibinfo {pages} {041601} (\bibinfo {year} {2006})}\BibitemShut {NoStop}%
\bibitem [{\citenamefont {Gou\"{e}t}\ \emph {et~al.}(2010)\citenamefont {Gou\"{e}t}, \citenamefont {Venkatraman}, \citenamefont {Wong},\ and\ \citenamefont {Shapiro}}]{LeGouet:10}%
  \BibitemOpen
  \bibfield  {author} {\bibinfo {author} {\bibfnamefont {J.~L.}\ \bibnamefont {Gou\"{e}t}}, \bibinfo {author} {\bibfnamefont {D.}~\bibnamefont {Venkatraman}}, \bibinfo {author} {\bibfnamefont {F.~N.~C.}\ \bibnamefont {Wong}}, \ and\ \bibinfo {author} {\bibfnamefont {J.~H.}\ \bibnamefont {Shapiro}},\ }\href {\doibase 10.1364/OL.35.001001} {\bibfield  {journal} {\bibinfo  {journal} {Opt. Lett.}\ }\textbf {\bibinfo {volume} {35}},\ \bibinfo {pages} {1001} (\bibinfo {year} {2010})}\BibitemShut {NoStop}%
\bibitem [{\citenamefont {Kaltenbaek}\ \emph {et~al.}(2009)\citenamefont {Kaltenbaek}, \citenamefont {Lavoie},\ and\ \citenamefont {Resch}}]{PhysRevLett.102.243601}%
  \BibitemOpen
  \bibfield  {author} {\bibinfo {author} {\bibfnamefont {R.}~\bibnamefont {Kaltenbaek}}, \bibinfo {author} {\bibfnamefont {J.}~\bibnamefont {Lavoie}}, \ and\ \bibinfo {author} {\bibfnamefont {K.~J.}\ \bibnamefont {Resch}},\ }\href@noop {} {\bibfield  {journal} {\bibinfo  {journal} {Phys. Rev. Lett.}\ }\textbf {\bibinfo {volume} {102}},\ \bibinfo {pages} {243601} (\bibinfo {year} {2009})}\BibitemShut {NoStop}%
\bibitem [{\citenamefont {Agrawal}(2013)}]{agrawal_nonlinear_2013}%
  \BibitemOpen
  \bibfield  {author} {\bibinfo {author} {\bibfnamefont {G.~P.}\ \bibnamefont {Agrawal}},\ }\href@noop {} {\emph {\bibinfo {title} {Nonlinear fiber optics}}},\ \bibinfo {edition} {fifth edition}\ ed.\ (\bibinfo  {publisher} {Elsevier/Academic Press},\ \bibinfo {address} {Amsterdam},\ \bibinfo {year} {2013})\BibitemShut {NoStop}%
\bibitem [{\citenamefont {Boto}\ \emph {et~al.}(2000)\citenamefont {Boto}, \citenamefont {Kok}, \citenamefont {Abrams}, \citenamefont {Braunstein}, \citenamefont {Williams},\ and\ \citenamefont {Dowling}}]{noon_state_PhysRevLett.85.2733}%
  \BibitemOpen
  \bibfield  {author} {\bibinfo {author} {\bibfnamefont {A.~N.}\ \bibnamefont {Boto}}, \bibinfo {author} {\bibfnamefont {P.}~\bibnamefont {Kok}}, \bibinfo {author} {\bibfnamefont {D.~S.}\ \bibnamefont {Abrams}}, \bibinfo {author} {\bibfnamefont {S.~L.}\ \bibnamefont {Braunstein}}, \bibinfo {author} {\bibfnamefont {C.~P.}\ \bibnamefont {Williams}}, \ and\ \bibinfo {author} {\bibfnamefont {J.~P.}\ \bibnamefont {Dowling}},\ }\href {\doibase 10.1103/PhysRevLett.85.2733} {\bibfield  {journal} {\bibinfo  {journal} {Phys. Rev. Lett.}\ }\textbf {\bibinfo {volume} {85}},\ \bibinfo {pages} {2733} (\bibinfo {year} {2000})}\BibitemShut {NoStop}%
\bibitem [{\citenamefont {Franson}(1989)}]{noon_time_PhysRevLett.62.2205}%
  \BibitemOpen
  \bibfield  {author} {\bibinfo {author} {\bibfnamefont {J.~D.}\ \bibnamefont {Franson}},\ }\href {\doibase 10.1103/PhysRevLett.62.2205} {\bibfield  {journal} {\bibinfo  {journal} {Phys. Rev. Lett.}\ }\textbf {\bibinfo {volume} {62}},\ \bibinfo {pages} {2205} (\bibinfo {year} {1989})}\BibitemShut {NoStop}%
\bibitem [{\citenamefont {Kim}\ \emph {et~al.}(2006)\citenamefont {Kim}, \citenamefont {Fiorentino},\ and\ \citenamefont {Wong}}]{noon_polar_PhysRevA.73.012316}%
  \BibitemOpen
  \bibfield  {author} {\bibinfo {author} {\bibfnamefont {T.}~\bibnamefont {Kim}}, \bibinfo {author} {\bibfnamefont {M.}~\bibnamefont {Fiorentino}}, \ and\ \bibinfo {author} {\bibfnamefont {F.~N.~C.}\ \bibnamefont {Wong}},\ }\href {\doibase 10.1103/PhysRevA.73.012316} {\bibfield  {journal} {\bibinfo  {journal} {Phys. Rev. A}\ }\textbf {\bibinfo {volume} {73}},\ \bibinfo {pages} {012316} (\bibinfo {year} {2006})}\BibitemShut {NoStop}%
\bibitem [{\citenamefont {Untern\"{a}hrer}\ \emph {et~al.}(2018)\citenamefont {Untern\"{a}hrer}, \citenamefont {Bessire}, \citenamefont {Gasparini}, \citenamefont {Perenzoni},\ and\ \citenamefont {Stefanov}}]{metro_1_Unternahrer:18}%
  \BibitemOpen
  \bibfield  {author} {\bibinfo {author} {\bibfnamefont {M.}~\bibnamefont {Untern\"{a}hrer}}, \bibinfo {author} {\bibfnamefont {B.}~\bibnamefont {Bessire}}, \bibinfo {author} {\bibfnamefont {L.}~\bibnamefont {Gasparini}}, \bibinfo {author} {\bibfnamefont {M.}~\bibnamefont {Perenzoni}}, \ and\ \bibinfo {author} {\bibfnamefont {A.}~\bibnamefont {Stefanov}},\ }\href {\doibase 10.1364/OPTICA.5.001150} {\bibfield  {journal} {\bibinfo  {journal} {Optica}\ }\textbf {\bibinfo {volume} {5}},\ \bibinfo {pages} {1150} (\bibinfo {year} {2018})}\BibitemShut {NoStop}%
\bibitem [{\citenamefont {S\'anchez Mu\~noz}\ \emph {et~al.}(2021)\citenamefont {S\'anchez Mu\~noz}, \citenamefont {Frascella},\ and\ \citenamefont {Schlawin}}]{metro_2_PhysRevResearch.3.033250}%
  \BibitemOpen
  \bibfield  {author} {\bibinfo {author} {\bibfnamefont {C.}~\bibnamefont {S\'anchez Mu\~noz}}, \bibinfo {author} {\bibfnamefont {G.}~\bibnamefont {Frascella}}, \ and\ \bibinfo {author} {\bibfnamefont {F.}~\bibnamefont {Schlawin}},\ }\href {\doibase 10.1103/PhysRevResearch.3.033250} {\bibfield  {journal} {\bibinfo  {journal} {Phys. Rev. Res.}\ }\textbf {\bibinfo {volume} {3}},\ \bibinfo {pages} {033250} (\bibinfo {year} {2021})}\BibitemShut {NoStop}%
\bibitem [{\citenamefont {Defienne}\ \emph {et~al.}(2022)\citenamefont {Defienne}, \citenamefont {Cameron}, \citenamefont {Ndagano}, \citenamefont {Lyons}, \citenamefont {Reichert}, \citenamefont {Zhao}, \citenamefont {Harvey}, \citenamefont {Charbon}, \citenamefont {Fleischer},\ and\ \citenamefont {Faccio}}]{metro_3_Defienne2022}%
  \BibitemOpen
  \bibfield  {author} {\bibinfo {author} {\bibfnamefont {H.}~\bibnamefont {Defienne}}, \bibinfo {author} {\bibfnamefont {P.}~\bibnamefont {Cameron}}, \bibinfo {author} {\bibfnamefont {B.}~\bibnamefont {Ndagano}}, \bibinfo {author} {\bibfnamefont {A.}~\bibnamefont {Lyons}}, \bibinfo {author} {\bibfnamefont {M.}~\bibnamefont {Reichert}}, \bibinfo {author} {\bibfnamefont {J.}~\bibnamefont {Zhao}}, \bibinfo {author} {\bibfnamefont {A.~R.}\ \bibnamefont {Harvey}}, \bibinfo {author} {\bibfnamefont {E.}~\bibnamefont {Charbon}}, \bibinfo {author} {\bibfnamefont {J.~W.}\ \bibnamefont {Fleischer}}, \ and\ \bibinfo {author} {\bibfnamefont {D.}~\bibnamefont {Faccio}},\ }\href {\doibase 10.1038/s41467-022-31052-6} {\bibfield  {journal} {\bibinfo  {journal} {Nature Communications}\ }\textbf {\bibinfo {volume} {13}},\ \bibinfo {pages} {3566} (\bibinfo {year} {2022})}\BibitemShut {NoStop}%
\bibitem [{\citenamefont {Yurke}\ \emph {et~al.}(1986)\citenamefont {Yurke}, \citenamefont {McCall},\ and\ \citenamefont {Klauder}}]{su11_PhysRevA.33.4033}%
  \BibitemOpen
  \bibfield  {author} {\bibinfo {author} {\bibfnamefont {B.}~\bibnamefont {Yurke}}, \bibinfo {author} {\bibfnamefont {S.~L.}\ \bibnamefont {McCall}}, \ and\ \bibinfo {author} {\bibfnamefont {J.~R.}\ \bibnamefont {Klauder}},\ }\href {\doibase 10.1103/PhysRevA.33.4033} {\bibfield  {journal} {\bibinfo  {journal} {Phys. Rev. A}\ }\textbf {\bibinfo {volume} {33}},\ \bibinfo {pages} {4033} (\bibinfo {year} {1986})}\BibitemShut {NoStop}%
\bibitem [{\citenamefont {Kaiser}\ \emph {et~al.}(2018)\citenamefont {Kaiser}, \citenamefont {Vergyris}, \citenamefont {Aktas}, \citenamefont {Babin}, \citenamefont {Labonté},\ and\ \citenamefont {Tanzilli}}]{kaiser_quantum_2018}%
  \BibitemOpen
  \bibfield  {author} {\bibinfo {author} {\bibfnamefont {F.}~\bibnamefont {Kaiser}}, \bibinfo {author} {\bibfnamefont {P.}~\bibnamefont {Vergyris}}, \bibinfo {author} {\bibfnamefont {D.}~\bibnamefont {Aktas}}, \bibinfo {author} {\bibfnamefont {C.}~\bibnamefont {Babin}}, \bibinfo {author} {\bibfnamefont {L.}~\bibnamefont {Labonté}}, \ and\ \bibinfo {author} {\bibfnamefont {S.}~\bibnamefont {Tanzilli}},\ }\href {\doibase 10.1038/lsa.2017.163} {\bibfield  {journal} {\bibinfo  {journal} {Light: Science \& Applications}\ }\textbf {\bibinfo {volume} {7}},\ \bibinfo {pages} {17163} (\bibinfo {year} {2018})}\BibitemShut {NoStop}%
\bibitem [{\citenamefont {Franson}(1992)}]{franson_disp_PhysRevA.45.3126}%
  \BibitemOpen
  \bibfield  {author} {\bibinfo {author} {\bibfnamefont {J.~D.}\ \bibnamefont {Franson}},\ }\href {\doibase 10.1103/PhysRevA.45.3126} {\bibfield  {journal} {\bibinfo  {journal} {Phys. Rev. A}\ }\textbf {\bibinfo {volume} {45}},\ \bibinfo {pages} {3126} (\bibinfo {year} {1992})}\BibitemShut {NoStop}%
\bibitem [{\citenamefont {Hlubina}\ \emph {et~al.}(2013)\citenamefont {Hlubina}, \citenamefont {Kadulová},\ and\ \citenamefont {Mergo}}]{hlubina_chromatic_2013}%
  \BibitemOpen
  \bibfield  {author} {\bibinfo {author} {\bibfnamefont {P.}~\bibnamefont {Hlubina}}, \bibinfo {author} {\bibfnamefont {M.}~\bibnamefont {Kadulová}}, \ and\ \bibinfo {author} {\bibfnamefont {P.}~\bibnamefont {Mergo}},\ }\href {\doibase 10.1016/j.optlaseng.2012.11.011} {\bibfield  {journal} {\bibinfo  {journal} {Optics and Lasers in Engineering}\ }\textbf {\bibinfo {volume} {51}},\ \bibinfo {pages} {421} (\bibinfo {year} {2013})}\BibitemShut {NoStop}%
\bibitem [{\citenamefont {Jachura}\ \emph {et~al.}(2014)\citenamefont {Jachura}, \citenamefont {Karpi?ski}, \citenamefont {Radzewicz},\ and\ \citenamefont {Banaszek}}]{jachura_high-visibility_2014}%
  \BibitemOpen
  \bibfield  {author} {\bibinfo {author} {\bibfnamefont {M.}~\bibnamefont {Jachura}}, \bibinfo {author} {\bibfnamefont {M.}~\bibnamefont {Karpi?ski}}, \bibinfo {author} {\bibfnamefont {C.}~\bibnamefont {Radzewicz}}, \ and\ \bibinfo {author} {\bibfnamefont {K.}~\bibnamefont {Banaszek}},\ }\href {\doibase 10.1364/OE.22.008624} {\bibfield  {journal} {\bibinfo  {journal} {Opt. Express}\ }\textbf {\bibinfo {volume} {22}},\ \bibinfo {pages} {8624} (\bibinfo {year} {2014})}\BibitemShut {NoStop}%
\bibitem [{\citenamefont {Labonté}\ \emph {et~al.}(2006)\citenamefont {Labonté}, \citenamefont {Roy}, \citenamefont {Pagnoux}, \citenamefont {Louradour}, \citenamefont {Restoin}, \citenamefont {Mélin},\ and\ \citenamefont {Burov}}]{labonte_experimental_2006}%
  \BibitemOpen
  \bibfield  {author} {\bibinfo {author} {\bibfnamefont {L.}~\bibnamefont {Labonté}}, \bibinfo {author} {\bibfnamefont {P.}~\bibnamefont {Roy}}, \bibinfo {author} {\bibfnamefont {D.}~\bibnamefont {Pagnoux}}, \bibinfo {author} {\bibfnamefont {F.}~\bibnamefont {Louradour}}, \bibinfo {author} {\bibfnamefont {C.}~\bibnamefont {Restoin}}, \bibinfo {author} {\bibfnamefont {G.}~\bibnamefont {Mélin}}, \ and\ \bibinfo {author} {\bibfnamefont {E.}~\bibnamefont {Burov}},\ }\href {\doibase 10.1088/1464-4258/8/11/001} {\bibfield  {journal} {\bibinfo  {journal} {J. Opt. A: Pure Appl. Opt.}\ }\textbf {\bibinfo {volume} {8}},\ \bibinfo {pages} {933} (\bibinfo {year} {2006})}\BibitemShut {NoStop}%
\bibitem [{\citenamefont {Grósz}\ \emph {et~al.}(2014)\citenamefont {Grósz}, \citenamefont {Kovács}, \citenamefont {Kiss},\ and\ \citenamefont {Szipőcs}}]{grosz_measurement_2014}%
  \BibitemOpen
  \bibfield  {author} {\bibinfo {author} {\bibfnamefont {T.}~\bibnamefont {Grósz}}, \bibinfo {author} {\bibfnamefont {A.~P.}\ \bibnamefont {Kovács}}, \bibinfo {author} {\bibfnamefont {M.}~\bibnamefont {Kiss}}, \ and\ \bibinfo {author} {\bibfnamefont {R.}~\bibnamefont {Szipőcs}},\ }\href {\doibase 10.1364/AO.53.001929} {\bibfield  {journal} {\bibinfo  {journal} {Applied Optics}\ }\textbf {\bibinfo {volume} {53}},\ \bibinfo {pages} {1929} (\bibinfo {year} {2014})}\BibitemShut {NoStop}%
\bibitem [{\citenamefont {Hlubina}\ \emph {et~al.}(2007)\citenamefont {Hlubina}, \citenamefont {Chlebus},\ and\ \citenamefont {Ciprian}}]{hlubina_differential_2007}%
  \BibitemOpen
  \bibfield  {author} {\bibinfo {author} {\bibfnamefont {P.}~\bibnamefont {Hlubina}}, \bibinfo {author} {\bibfnamefont {R.}~\bibnamefont {Chlebus}}, \ and\ \bibinfo {author} {\bibfnamefont {D.}~\bibnamefont {Ciprian}},\ }\href {\doibase 10.1088/0957-0233/18/5/046} {\bibfield  {journal} {\bibinfo  {journal} {Measurement Science and Technology}\ }\textbf {\bibinfo {volume} {18}},\ \bibinfo {pages} {1547} (\bibinfo {year} {2007})}\BibitemShut {NoStop}%
\bibitem [{\citenamefont {Galle}\ \emph {et~al.}(2007)\citenamefont {Galle}, \citenamefont {Mohammed}, \citenamefont {Qian},\ and\ \citenamefont {Smith}}]{galle_single-arm_2007}%
  \BibitemOpen
  \bibfield  {author} {\bibinfo {author} {\bibfnamefont {M.~A.}\ \bibnamefont {Galle}}, \bibinfo {author} {\bibfnamefont {W.}~\bibnamefont {Mohammed}}, \bibinfo {author} {\bibfnamefont {L.}~\bibnamefont {Qian}}, \ and\ \bibinfo {author} {\bibfnamefont {P.~W.~E.}\ \bibnamefont {Smith}},\ }\href {\doibase 10.1364/OE.15.016896} {\bibfield  {journal} {\bibinfo  {journal} {Opt. Express, OE}\ }\textbf {\bibinfo {volume} {15}},\ \bibinfo {pages} {16896} (\bibinfo {year} {2007})}\BibitemShut {NoStop}%
\bibitem [{\citenamefont {Ye}\ \emph {et~al.}(2002)\citenamefont {Ye}, \citenamefont {Xu}, \citenamefont {Liu}, \citenamefont {Knox}, \citenamefont {Yan}, \citenamefont {Windeler},\ and\ \citenamefont {Eggleton}}]{ye_dispersion_2002}%
  \BibitemOpen
  \bibfield  {author} {\bibinfo {author} {\bibfnamefont {Q.}~\bibnamefont {Ye}}, \bibinfo {author} {\bibfnamefont {C.}~\bibnamefont {Xu}}, \bibinfo {author} {\bibfnamefont {X.}~\bibnamefont {Liu}}, \bibinfo {author} {\bibfnamefont {W.~H.}\ \bibnamefont {Knox}}, \bibinfo {author} {\bibfnamefont {M.~F.}\ \bibnamefont {Yan}}, \bibinfo {author} {\bibfnamefont {R.~S.}\ \bibnamefont {Windeler}}, \ and\ \bibinfo {author} {\bibfnamefont {B.}~\bibnamefont {Eggleton}},\ }\href {\doibase 10.1364/AO.41.004467} {\bibfield  {journal} {\bibinfo  {journal} {Appl. Opt., AO}\ }\textbf {\bibinfo {volume} {41}},\ \bibinfo {pages} {4467} (\bibinfo {year} {2002})}\BibitemShut {NoStop}%
\bibitem [{\citenamefont {Qin}\ \emph {et~al.}(2023)\citenamefont {Qin}, \citenamefont {Deng}, \citenamefont {Zhong}, \citenamefont {Peng}, \citenamefont {Su}, \citenamefont {Luo}, \citenamefont {Xu}, \citenamefont {Wu}, \citenamefont {Gong}, \citenamefont {Liu}, \citenamefont {Wang}, \citenamefont {Chen}, \citenamefont {Li}, \citenamefont {Liu}, \citenamefont {Lu},\ and\ \citenamefont {Pan}}]{qin_unconditional_2023}%
  \BibitemOpen
  \bibfield  {author} {\bibinfo {author} {\bibfnamefont {J.}~\bibnamefont {Qin}}, \bibinfo {author} {\bibfnamefont {Y.-H.}\ \bibnamefont {Deng}}, \bibinfo {author} {\bibfnamefont {H.-S.}\ \bibnamefont {Zhong}}, \bibinfo {author} {\bibfnamefont {L.-C.}\ \bibnamefont {Peng}}, \bibinfo {author} {\bibfnamefont {H.}~\bibnamefont {Su}}, \bibinfo {author} {\bibfnamefont {Y.-H.}\ \bibnamefont {Luo}}, \bibinfo {author} {\bibfnamefont {J.-M.}\ \bibnamefont {Xu}}, \bibinfo {author} {\bibfnamefont {D.}~\bibnamefont {Wu}}, \bibinfo {author} {\bibfnamefont {S.-Q.}\ \bibnamefont {Gong}}, \bibinfo {author} {\bibfnamefont {H.-L.}\ \bibnamefont {Liu}}, \bibinfo {author} {\bibfnamefont {H.}~\bibnamefont {Wang}}, \bibinfo {author} {\bibfnamefont {M.-C.}\ \bibnamefont {Chen}}, \bibinfo {author} {\bibfnamefont {L.}~\bibnamefont {Li}}, \bibinfo {author} {\bibfnamefont {N.-L.}\ \bibnamefont {Liu}}, \bibinfo {author} {\bibfnamefont {C.-Y.}\ \bibnamefont {Lu}}, \ and\ \bibinfo {author} {\bibfnamefont {J.-W.}\ \bibnamefont {Pan}},\
  }\href {\doibase 10.1103/PhysRevLett.130.070801} {\bibfield  {journal} {\bibinfo  {journal} {Phys. Rev. Lett.}\ }\textbf {\bibinfo {volume} {130}},\ \bibinfo {pages} {070801} (\bibinfo {year} {2023})},\ \bibinfo {note} {publisher: American Physical Society}\BibitemShut {NoStop}%
\bibitem [{\citenamefont {Nold}\ \emph {et~al.}(2022)\citenamefont {Nold}, \citenamefont {Babin}, \citenamefont {Schmidt}, \citenamefont {Linkewitz}, \citenamefont {Zaballos}, \citenamefont {Stöhr}, \citenamefont {Kolesov}, \citenamefont {Vorobyov}, \citenamefont {Lukin}, \citenamefont {Boppert}, \citenamefont {Barz}, \citenamefont {Vučković}, \citenamefont {Gebhardt}, \citenamefont {Kaiser},\ and\ \citenamefont {Wrachtrup}}]{nold_quantum_2022}%
  \BibitemOpen
  \bibfield  {author} {\bibinfo {author} {\bibfnamefont {R.}~\bibnamefont {Nold}}, \bibinfo {author} {\bibfnamefont {C.}~\bibnamefont {Babin}}, \bibinfo {author} {\bibfnamefont {J.}~\bibnamefont {Schmidt}}, \bibinfo {author} {\bibfnamefont {T.}~\bibnamefont {Linkewitz}}, \bibinfo {author} {\bibfnamefont {M.~T.~P.}\ \bibnamefont {Zaballos}}, \bibinfo {author} {\bibfnamefont {R.}~\bibnamefont {Stöhr}}, \bibinfo {author} {\bibfnamefont {R.}~\bibnamefont {Kolesov}}, \bibinfo {author} {\bibfnamefont {V.}~\bibnamefont {Vorobyov}}, \bibinfo {author} {\bibfnamefont {D.~M.}\ \bibnamefont {Lukin}}, \bibinfo {author} {\bibfnamefont {R.}~\bibnamefont {Boppert}}, \bibinfo {author} {\bibfnamefont {S.}~\bibnamefont {Barz}}, \bibinfo {author} {\bibfnamefont {J.}~\bibnamefont {Vučković}}, \bibinfo {author} {\bibfnamefont {J.~C.~M.}\ \bibnamefont {Gebhardt}}, \bibinfo {author} {\bibfnamefont {F.}~\bibnamefont {Kaiser}}, \ and\ \bibinfo {author} {\bibfnamefont {J.}~\bibnamefont {Wrachtrup}},\ }\href {\doibase
  10.1103/PRXQuantum.3.020358} {\bibfield  {journal} {\bibinfo  {journal} {PRX Quantum}\ }\textbf {\bibinfo {volume} {3}},\ \bibinfo {pages} {020358} (\bibinfo {year} {2022})},\ \bibinfo {note} {publisher: American Physical Society}\BibitemShut {NoStop}%
\bibitem [{\citenamefont {Dorner}\ \emph {et~al.}(2009)\citenamefont {Dorner}, \citenamefont {Demkowicz-Dobrzanski}, \citenamefont {Smith}, \citenamefont {Lundeen}, \citenamefont {Wasilewski}, \citenamefont {Banaszek},\ and\ \citenamefont {Walmsley}}]{dorner_optimal_2009}%
  \BibitemOpen
  \bibfield  {author} {\bibinfo {author} {\bibfnamefont {U.}~\bibnamefont {Dorner}}, \bibinfo {author} {\bibfnamefont {R.}~\bibnamefont {Demkowicz-Dobrzanski}}, \bibinfo {author} {\bibfnamefont {B.~J.}\ \bibnamefont {Smith}}, \bibinfo {author} {\bibfnamefont {J.~S.}\ \bibnamefont {Lundeen}}, \bibinfo {author} {\bibfnamefont {W.}~\bibnamefont {Wasilewski}}, \bibinfo {author} {\bibfnamefont {K.}~\bibnamefont {Banaszek}}, \ and\ \bibinfo {author} {\bibfnamefont {I.~A.}\ \bibnamefont {Walmsley}},\ }\href {\doibase 10.1103/PhysRevLett.102.040403} {\bibfield  {journal} {\bibinfo  {journal} {Phys. Rev. Lett.}\ }\textbf {\bibinfo {volume} {102}},\ \bibinfo {pages} {040403} (\bibinfo {year} {2009})}\BibitemShut {NoStop}%
\bibitem [{\citenamefont {Demkowicz-Dobrzanski}\ \emph {et~al.}(2009)\citenamefont {Demkowicz-Dobrzanski}, \citenamefont {Dorner}, \citenamefont {Smith}, \citenamefont {Lundeen}, \citenamefont {Wasilewski}, \citenamefont {Banaszek},\ and\ \citenamefont {Walmsley}}]{demkowicz-dobrzanski_quantum_2009}%
  \BibitemOpen
  \bibfield  {author} {\bibinfo {author} {\bibfnamefont {R.}~\bibnamefont {Demkowicz-Dobrzanski}}, \bibinfo {author} {\bibfnamefont {U.}~\bibnamefont {Dorner}}, \bibinfo {author} {\bibfnamefont {B.~J.}\ \bibnamefont {Smith}}, \bibinfo {author} {\bibfnamefont {J.~S.}\ \bibnamefont {Lundeen}}, \bibinfo {author} {\bibfnamefont {W.}~\bibnamefont {Wasilewski}}, \bibinfo {author} {\bibfnamefont {K.}~\bibnamefont {Banaszek}}, \ and\ \bibinfo {author} {\bibfnamefont {I.~A.}\ \bibnamefont {Walmsley}},\ }\href {\doibase 10.1103/PhysRevA.80.013825} {\bibfield  {journal} {\bibinfo  {journal} {Phys. Rev. A}\ }\textbf {\bibinfo {volume} {80}},\ \bibinfo {pages} {013825} (\bibinfo {year} {2009})}\BibitemShut {NoStop}%
\bibitem [{\citenamefont {Kacprowicz}\ \emph {et~al.}(2010)\citenamefont {Kacprowicz}, \citenamefont {Demkowicz-Dobrzański}, \citenamefont {Wasilewski}, \citenamefont {Banaszek},\ and\ \citenamefont {Walmsley}}]{kacprowicz_experimental_2010}%
  \BibitemOpen
  \bibfield  {author} {\bibinfo {author} {\bibfnamefont {M.}~\bibnamefont {Kacprowicz}}, \bibinfo {author} {\bibfnamefont {R.}~\bibnamefont {Demkowicz-Dobrzański}}, \bibinfo {author} {\bibfnamefont {W.}~\bibnamefont {Wasilewski}}, \bibinfo {author} {\bibfnamefont {K.}~\bibnamefont {Banaszek}}, \ and\ \bibinfo {author} {\bibfnamefont {I.~A.}\ \bibnamefont {Walmsley}},\ }\href {\doibase 10.1038/nphoton.2010.39} {\bibfield  {journal} {\bibinfo  {journal} {Nature Photon}\ }\textbf {\bibinfo {volume} {4}},\ \bibinfo {pages} {357} (\bibinfo {year} {2010})}\BibitemShut {NoStop}%
\bibitem [{\citenamefont {Datta}\ \emph {et~al.}(2011)\citenamefont {Datta}, \citenamefont {Zhang}, \citenamefont {Thomas-Peter}, \citenamefont {Dorner}, \citenamefont {Smith},\ and\ \citenamefont {Walmsley}}]{datta_quantum_2011}%
  \BibitemOpen
  \bibfield  {author} {\bibinfo {author} {\bibfnamefont {A.}~\bibnamefont {Datta}}, \bibinfo {author} {\bibfnamefont {L.}~\bibnamefont {Zhang}}, \bibinfo {author} {\bibfnamefont {N.}~\bibnamefont {Thomas-Peter}}, \bibinfo {author} {\bibfnamefont {U.}~\bibnamefont {Dorner}}, \bibinfo {author} {\bibfnamefont {B.~J.}\ \bibnamefont {Smith}}, \ and\ \bibinfo {author} {\bibfnamefont {I.~A.}\ \bibnamefont {Walmsley}},\ }\href {\doibase 10.1103/PhysRevA.83.063836} {\bibfield  {journal} {\bibinfo  {journal} {Phys. Rev. A}\ }\textbf {\bibinfo {volume} {83}},\ \bibinfo {pages} {063836} (\bibinfo {year} {2011})}\BibitemShut {NoStop}%
\bibitem [{\citenamefont {Escher}\ \emph {et~al.}(2011)\citenamefont {Escher}, \citenamefont {De~Matos~Filho},\ and\ \citenamefont {Davidovich}}]{escher_general_2011}%
  \BibitemOpen
  \bibfield  {author} {\bibinfo {author} {\bibfnamefont {B.~M.}\ \bibnamefont {Escher}}, \bibinfo {author} {\bibfnamefont {R.~L.}\ \bibnamefont {De~Matos~Filho}}, \ and\ \bibinfo {author} {\bibfnamefont {L.}~\bibnamefont {Davidovich}},\ }\href {\doibase 10.1038/nphys1958} {\bibfield  {journal} {\bibinfo  {journal} {Nature Phys}\ }\textbf {\bibinfo {volume} {7}},\ \bibinfo {pages} {406} (\bibinfo {year} {2011})}\BibitemShut {NoStop}%
\bibitem [{\citenamefont {Thomas-Peter}\ \emph {et~al.}(2011)\citenamefont {Thomas-Peter}, \citenamefont {Smith}, \citenamefont {Datta}, \citenamefont {Zhang}, \citenamefont {Dorner},\ and\ \citenamefont {Walmsley}}]{thomas-peter_real-world_2011}%
  \BibitemOpen
  \bibfield  {author} {\bibinfo {author} {\bibfnamefont {N.}~\bibnamefont {Thomas-Peter}}, \bibinfo {author} {\bibfnamefont {B.~J.}\ \bibnamefont {Smith}}, \bibinfo {author} {\bibfnamefont {A.}~\bibnamefont {Datta}}, \bibinfo {author} {\bibfnamefont {L.}~\bibnamefont {Zhang}}, \bibinfo {author} {\bibfnamefont {U.}~\bibnamefont {Dorner}}, \ and\ \bibinfo {author} {\bibfnamefont {I.~A.}\ \bibnamefont {Walmsley}},\ }\href {\doibase 10.1103/PhysRevLett.107.113603} {\bibfield  {journal} {\bibinfo  {journal} {Phys. Rev. Lett.}\ }\textbf {\bibinfo {volume} {107}},\ \bibinfo {pages} {113603} (\bibinfo {year} {2011})}\BibitemShut {NoStop}%
\bibitem [{\citenamefont {Lemos}\ \emph {et~al.}(2014)\citenamefont {Lemos}, \citenamefont {Borish}, \citenamefont {Cole}, \citenamefont {Ramelow}, \citenamefont {Lapkiewicz},\ and\ \citenamefont {Zeilinger}}]{lemos_quantum_2014}%
  \BibitemOpen
  \bibfield  {author} {\bibinfo {author} {\bibfnamefont {G.~B.}\ \bibnamefont {Lemos}}, \bibinfo {author} {\bibfnamefont {V.}~\bibnamefont {Borish}}, \bibinfo {author} {\bibfnamefont {G.~D.}\ \bibnamefont {Cole}}, \bibinfo {author} {\bibfnamefont {S.}~\bibnamefont {Ramelow}}, \bibinfo {author} {\bibfnamefont {R.}~\bibnamefont {Lapkiewicz}}, \ and\ \bibinfo {author} {\bibfnamefont {A.}~\bibnamefont {Zeilinger}},\ }\href {\doibase 10.1038/nature13586} {\bibfield  {journal} {\bibinfo  {journal} {Nature}\ }\textbf {\bibinfo {volume} {512}},\ \bibinfo {pages} {409} (\bibinfo {year} {2014})}\BibitemShut {NoStop}%
\bibitem [{\citenamefont {Crespi}\ \emph {et~al.}(2012)\citenamefont {Crespi}, \citenamefont {Lobino}, \citenamefont {Matthews}, \citenamefont {Politi}, \citenamefont {Neal}, \citenamefont {Ramponi}, \citenamefont {Osellame},\ and\ \citenamefont {O’Brien}}]{crespi_measuring_2012}%
  \BibitemOpen
  \bibfield  {author} {\bibinfo {author} {\bibfnamefont {A.}~\bibnamefont {Crespi}}, \bibinfo {author} {\bibfnamefont {M.}~\bibnamefont {Lobino}}, \bibinfo {author} {\bibfnamefont {J.~C.~F.}\ \bibnamefont {Matthews}}, \bibinfo {author} {\bibfnamefont {A.}~\bibnamefont {Politi}}, \bibinfo {author} {\bibfnamefont {C.~R.}\ \bibnamefont {Neal}}, \bibinfo {author} {\bibfnamefont {R.}~\bibnamefont {Ramponi}}, \bibinfo {author} {\bibfnamefont {R.}~\bibnamefont {Osellame}}, \ and\ \bibinfo {author} {\bibfnamefont {J.~L.}\ \bibnamefont {O’Brien}},\ }\href {\doibase 10.1063/1.4724105} {\bibfield  {journal} {\bibinfo  {journal} {Applied Physics Letters}\ }\textbf {\bibinfo {volume} {100}},\ \bibinfo {pages} {233704} (\bibinfo {year} {2012})}\BibitemShut {NoStop}%
\bibitem [{\citenamefont {Chen}\ \emph {et~al.}(2022)\citenamefont {Chen}, \citenamefont {Hong},\ and\ \citenamefont {Chen}}]{chen_quantum_2022}%
  \BibitemOpen
  \bibfield  {author} {\bibinfo {author} {\bibfnamefont {Y.}~\bibnamefont {Chen}}, \bibinfo {author} {\bibfnamefont {L.}~\bibnamefont {Hong}}, \ and\ \bibinfo {author} {\bibfnamefont {L.}~\bibnamefont {Chen}},\ }\href {\doibase 10.3389/fphy.2022.892519} {\bibfield  {journal} {\bibinfo  {journal} {Front. Phys.}\ }\textbf {\bibinfo {volume} {10}},\ \bibinfo {pages} {892519} (\bibinfo {year} {2022})}\BibitemShut {NoStop}%
\end{thebibliography}

%

\end{document}